\documentclass[twocolumn]{aastex62}
\usepackage{amsfonts,amsmath,amssymb,amsthm,epsfig,graphicx,float,tabularx,enumitem}
\begin{document}

\title{Tidal migration of exoplanets around M-dwarfs: frequency-dependent tidal dissipation}

\correspondingauthor{Samantha Wu}
\email{scwu@astro.caltech.edu }

\author{Samantha C. Wu}
\affiliation{California Institute of Technology, Astronomy Department, Pasadena, CA 91125, USA}
\affiliation{TAPIR, Mailcode 350-17, California Institute of Technology, Pasadena, CA 91125, USA}
\author{Janosz W. Dewberry}
\affiliation{CITA, 60 St. George Street, Toronto, ON M5S 3H8, Canada}

\author{Jim Fuller}
\affiliation{TAPIR, Mailcode 350-17, California Institute of Technology, Pasadena, CA 91125, USA}

\begin{abstract}
The orbital architectures of short-period exoplanet systems are shaped by tidal dissipation in their host stars. For low-mass M-dwarfs whose dynamical tidal response comprises a dense spectrum of inertial modes at low frequencies, resolving the frequency dependence of tidal dissipation is crucial to capturing the effect of tides on planetary orbits throughout the evolutionary stages of the host star. We use non-perturbative spectral methods to calculate the normal mode oscillations of a fully-convective M-dwarf modeled using realistic stellar profiles from MESA. We compute the dissipative tidal response composed of contributions from each mode as well as non-adiabatic coupling between the modes, which we find to be an essential component of the dissipative calculations. 
% Using our results for the tidal dissipation across the pre-main sequence to main-sequence evolution of the star, 
% By performing integrations of the evolution of circular, coplanar planetary orbits under the influence of tides, we examine the effect of tidal dissipation in the host star upon the orbital separations of its planetary companions.
Using our results for dissipation, we then compute the evolution of circular, coplanar planetary orbits under the influence of tides in the host star.
We find that orbital migration driven by resonance locking affects the orbits of Earth-mass planets at orbital periods $P_{\rm orb} \lesssim 1.5$ day and of Jupiter-mass planets at $P_{\rm orb} \lesssim 2.5$ day.  Due to resonantly-driven orbital decay and outward migration, we predict a dearth of small planets closer than $P_{\rm orb} \sim 1$ day and similarly sparse numbers of more massive planets out to $P_{\rm orb} \sim 3$ day.
%should we mention varying viscosity?
\end{abstract}

\section{Introduction}

Through missions such as Kepler and TESS, thousands of exoplanets with a variety of sizes, masses, and orbital architectures have been discovered to date around stars of spectral types M through A. Due to the observational bias of both the transit and radial velocity methods, a large proportion of exoplanets are detected at short orbital periods, where their proximity to the host star can shape these planetary orbits through tidal effects. Tidal interactions may have modified the orbital inclinations, eccentricities, and semimajor axes across the history of these planetary systems. 

In at least one case, tides are suspected to be acting in real time, as the orbit of exoplanet WASP-12b is measurably decaying from its current orbital period of $P_{\rm orb} = 1.09$ day according to transit-timing analysis over the past decade \citep{maciejewski2013,maciejewski2018,patra2017,yee2020}. Due to the short decay timescale of $\sim$3 Myr, it is thought that the influence of the so-called dynamical tide in the late-F star host is accelerating the infall \citep{weinberg2017,bailey2019,barker2020}. In addition to the equilibrium tide \citep{zahn1966a}, which represents the hydrostatic response of the star to its exoplanet perturber, the dynamical tide in the star is the oscillatory component of the response. 
% Tidal dissipation within the star occurs as either the quasi-hydrostatic distortion of the equilibrium tide or 
The oscillation modes of the dynamical tide are damped due to local effects, such as turbulent viscosity in convective zones or radiative damping in stably-stratified stellar regions. This process leads to an exchange of energy and angular momentum between the orbit and the stellar interior.
%introduce exoplanet research, importance of tides in the observations
% Bring up the WASP 12b whose tidal decay is measured (see https://iopscience.iop.org/article/10.3847/2041-8213/ab8563 for the references) with Q of about 2e5, decay timescale \sim 3  Myr
% https://ui.adsabs.harvard.edu/abs/2023MNRAS.525..876M/abstract: close-in giant planets around partially convective stars have Q~1e5.

As the majority of stars in the galaxy are low-mass, the characteristics of M-dwarfs as planetary hosts are important for understanding the effects of tides in the most common exoplanet systems. Though the low luminosity of M-dwarfs makes them intrinsically more difficult to observe, short-period exoplanet systems around M-dwarfs are of interest as they are likely to be located in the habitable zone of the M-dwarfs \citep[e.g., Trappist-1,][]{2017Natur.542..456G} and are relatively easy to detect. Statistical analysis of exoplanet detections indicates that close-in exoplanets around M-dwarfs are much more abundant than around other spectral types, and they are also more likely to be small with $R \lesssim R_{\oplus}$ \citep{Mulders_2015,hsu2020,dressing2013,ment2023}. With TESS already contributing dozens of exoplanets with measured masses and radii hosted by M-dwarfs and the Nancy Grace Roman Space Telescope anticipated to yield $\sim \! 10^3$ small planets around early- and mid-M dwarfs \citep{tamburo2023}, the plethora of observed systems will facilitate tests of theoretical predictions for how planetary architectures are shaped by effects such as tides. 
%narrow in on M dwarfs and how they are common hosts which are likely to host habitable, earth-like planets e.g. Trappist I. and many more will be found as the surveys are already getting better (e.g. TESS)
% Why study Mdwarfs? Habitable zone, higher occurrence rate of close-in planets
% Bring up Trappist system
% Bring up huge amount of planets around M dwarfs discovered by TESS recently - dozens with measured masses and radii
%"most, and potentially all, early-type M dwarfs harbour planetary systems." Hsu; also note Mulders and Hsu that both host more close-in planets that are Earth mass. Ment and Charbonneau also find that for M dwarfs between 0.1-0.3 Msun in TESS the occurrence rate is 0.6 planets per star between 0.4-7 day note that jupiter mass is less expected bc disks are not massive, but massive jupiters have been found (cite - see ads library)

Studies of tidal dissipation in low-mass stellar hosts have accounted for the influential contribution of the dynamical tide, which can provide orders of magnitude larger dissipation than the equilibrium tide \citep{ogilvie2007,bolmont2016}. Many also follow how the strength of the tidal dissipation varies during the evolution of the star. In these low-mass stars ($M \lesssim 0.35\, M_{\odot}$) that are fully convective during the pre-main sequence and the main sequence for $\sim \! 10^{11\text{--}12}$ yr, inertial modes (i-modes) excited in the stellar convective zone comprise the dynamical tide. Restored by the Coriolis force, i-modes propagate with rotating-frame frequencies $-2\Omega_s < \omega < 2\Omega_s$, where $\Omega_s$ is the stellar spin frequency, and may be damped by a convective turbulent viscosity. Stars containing radiative zones also experience dissipation due to radiative damping of internal gravity waves. At larger frequencies, the contribution from fundamental modes (f-modes), also described as surface gravity modes, constitutes the majority of the star's equilibrium tidal response. 

Previous works have parameterized the strength of the dynamical tide due to i-modes as a function of the stellar properties in order to study the tidal dissipation across stellar evolution. In the simplest case of i-modes excited in a homogeneous envelope surrounding a rigid core with fractional radius $\alpha$, averaging over the i-mode frequency range $-2\Omega_s < \omega < 2\Omega_s$ produces a dissipation rate that scales as $\alpha^5$ and the dimensionless stellar rotation rate $\epsilon^2$ \citep{goodman2009,ogilvie2013}. With this formula for the dynamical tidal dissipation as a function of stellar properties, many studies have efficiently calculated the tidal dissipation coupled with stellar evolution \citep{mathis2016,bolmont2016,gallet2017,bolmont2017}. However, the formula that is most commonly used in these works is of limited use for fully convective stars with realistic density profiles, as it predicts negligible tidal dissipation as the core size $\alpha\rightarrow 0$. In contrast, tidally-excited inertial modes have actually been demonstrated to contribute significantly to the dynamical tide in rapidly-rotating, coreless isentropic bodies \citep{wu2005a,wu2005b,ogilvie2013,Dewberry2022,Dewberry2023}.
% \cite{barker2020,Barker_2022} use a more realistic density profile

Furthermore, studies employing a frequency-averaged formalism do not account for the strong frequency dependence of the dynamical tidal response, which exhibits large peaks associated with resonances with the inertial modes \citep{ogilvie2007,ogilvie2013}. 
Ignoring the frequency dependence of tidal dissipation also entails the drawback that no resonance locking can be resolved in these systems. During resonance locking, the enhanced response resulting from tidal excitation of resonant modes persists over a significant period of the stellar lifetime. This mechanism has therefore been invoked to explain faster than expected orbital migration and circularization in stellar binaries and planet-moon systems \citep{witte1999,fuller2012,fuller2016}. In order to capture the amplified response near mode resonances, the tidal dissipation as a function of frequency must be calculated throughout the stellar evolution. For convective stars, resonance locking is of particular interest since the frequency range of i-modes coincides with the tidal forcing frequencies of most planetary companions. Thus, planetary orbits will have many opportunities to encounter resonances with the capacity to shape orbital architectures through strong tidal dissipation.  

%introduce previous work on tidal dissipation in exoplanet systems, but note that the freq avg dissipation is commonly used and that it predicts zero dynamical tide for a fully convective star, so fully convective stars have been previously ignored.
% Discuss previous efforts studying dissipation around M dwarfs (Gallet, Bolmont, Mathis works) 
% 	see dissipation from freq-ind dissipation. No res locks seen because it doesn't depend on frequency
% 	Don't find dissipation in fully convective stars (show eq'n to explain why)
% Note drawbacks of this approach and what can be done instead
% 	Show Janosz's work, see the modes
% 	Especially i-modes in isentropic polytropes from Janosz's paper - demonstrates that there can be dissipation in an isentropic star
%   Also note that i-modes are in the frequency range of typical orbital periods

In this work, we perform frequency-dependent calculations of tidal dissipation, coupled with the stellar evolution of a $M=0.2\, M_{\odot}$ M-dwarf which is fully convective on the pre-main sequence (PMS) and main sequence (MS). This procedure is repeated for models with different initial rotation periods, allowing us to capture the diversity in outcomes that ensue by varying this initial condition. Using an expansion over the normal modes of stellar oscillation, we compute the dissipative response of the star, thereby resolving the resonant frequencies coinciding with inertial mode (i-mode) and fundamental mode (f-mode) frequencies as a function of stellar age. We use our results to calculate the orbital migration of Earth-mass and Jupiter-mass planets, initialized at a range of ages to represent a few possible formation pathways, and find that resonance locking dominates the migration of planets within $\sim 1$--$2$ day orbits. Section \ref{sec:physics} describes how we set up our calculations of the tidal dissipation, normal mode eigenfunctions, and orbital evolution; Section \ref{sec:dissipation_results} demonstrates the dissipative response of the star across its evolution, noting how our approach compares to other methods; and Section \ref{sec:orbital_results} shows the associated orbital migration due to tides. We discuss the observational implications for exoplanet demographics and the uncertainties in our implementation of the tidal response in Section \ref{sec:discussion}.

\section{Methods}
\label{sec:physics}

\subsection{Tidal Formalism}
\label{sec:tides}
We consider circular, coplanar orbits of a satellite of mass $M'$ at a separation $a$ from a body of mass $M$, which rotates at a spin frequency $\Omega_s$. The satellite orbital frequency is $\Omega_o = [G(M+M')/a^3]^{1/2}$.
In a frame centered on and rotating with the body of mass $M$ which experiences the tidal disturbance, the tidal potential of the satellite can be written as
% \begin{equation}
%     U=-\frac{GM'}{a}
%     \sum_{\ell=2}^\infty\sum_{m=-l}^l
%     \frac{4\pi}{2\ell+1}
%     \left(\frac{r}{a}\right)^l
%     Y_l^{m*}(\theta',\phi')Y_l^m(\theta,\phi),
% \end{equation}
\begin{align}
    U=\sum_{\ell=2}^\infty\sum_{m=-\ell}^\ell
    U_{\ell m}r^\ell Y_\ell^m(\theta,\phi)
    \exp[-\text{i}\omega_mt],
\end{align}
where
\begin{align}
    \omega_m&=m(\Omega_o-\Omega_s),
\\
    U_{\ell m}&=-\left(\frac{GM'}{a^{l+1}}\right) \left(\frac{4\pi}{2\ell+1}\right)Y_{\ell}^{m*}(\pi/2,0).
    %W_{\ell m},
%\\
    %W_{\ell m}&=
    % \left(\frac{4\pi}{2\ell+1}\right)Y_{\ell}^{m*}(\pi/2,0)
% \\\notag
%     &=(-1)^{(\ell+m)/2}
%     \left[ 
%         \frac{4\pi(\ell+m)!(\ell-m)!}
%         {2\ell + 1}
%     \right]^{1/2}
% \\\notag
%     &     \times \left[ 
%         2^\ell
%         \left( 
%             \frac{\ell+m}{2}
%         \right)! 
%         \left( 
%             \frac{\ell-m}{2}
%         \right)! 
%     \right]^{-1},
\end{align}
% and $W_{\ell m}=0$ for odd $\ell+m$.

From here onward, we refer to the perturbed body $M$ as a star, and the satellite $M'$ as a planet.
The linearized momentum equation for Lagrangian displacement $\boldsymbol{\xi}$ in the star's rotating frame of reference is 
\begin{equation}\label{eq:EOM}
    \frac{\partial^2\boldsymbol{\xi}}{\partial t^2}
    +2{\bf \Omega_s\times}\frac{\partial \boldsymbol{\xi}}{\partial t}
    +{\bf C}[\boldsymbol{\xi}]
    =-\nabla U,
\end{equation}
where ${\bf C}$ is a self-adjoint operator describing internal fluid forces.
The normal modes of the star are $\hat{\boldsymbol{\xi}}_\alpha= \boldsymbol{\xi}_\alpha({\bf r})\exp[-\text{i}\omega_\alpha t]$. For a rotating fluid, the normal modes satisfy the following conditions (in units of $G=M=R=1$, where $M$ and $R$ are the stellar mass and radius respectively):
\begin{align}
\label{eq:EoM}
    -\omega_\alpha^2\boldsymbol{\xi}_\beta
    -2\mathrm{i}\omega_\alpha{\bf \Omega_s\times}\boldsymbol{\xi}_\alpha
    +{\textbf C}[\boldsymbol{\xi}_\alpha] &=0,
\\
\label{eq:orthogonality}
    (\omega_\alpha + \omega_\beta)
    \langle \boldsymbol{\xi}_\alpha,\boldsymbol{\xi}_\beta\rangle
    +\langle \boldsymbol{\xi}_\alpha,
    2\mathrm{i}{\bf \Omega\times}\boldsymbol{\xi}_\beta \rangle &=0,\, \mathrm{for}\, \beta\not=\alpha,
\\
    \langle \boldsymbol{\xi}_\alpha,\boldsymbol{\xi}_{\alpha}\rangle&=1, %MR^2,
\end{align}
where $\langle\,,\rangle$ is the inner product such that
\begin{equation}
     \langle 
        \boldsymbol{\xi}_\alpha,\boldsymbol{\xi}_\beta
    \rangle
    =\int_V\rho_0
    \boldsymbol{\xi}_\alpha^*\cdot\boldsymbol{\xi}_\beta\text{d}V.
\end{equation}
%if shortened or if needed could mention that $\xi = \sum_{\alpha} c_{\alpha}\xi_{\alpha}$
Assuming a phase space expansion of the tidal response in terms of normal modes of stellar oscillation, it can be shown (e.g. \citealt{Schenk2001,Lai2006}) that the amplitude of oscillation mode $\alpha$ forced by the tidal potential in the absence of dissipation satisfies
\begin{equation}
    \dot{c}_{\alpha}
    +\text{i}\omega_{\alpha}c_{\alpha}
    =-\frac{\text{i}}{2\epsilon_{\alpha}} \exp[-\text{i}\omega_mt]
    \sum_{\ell} U_{\ell m}Q_{\ell m}^\alpha,
\end{equation}
where $\epsilon_{\alpha}
=\omega_\alpha
\langle \boldsymbol{\xi}_\alpha,\boldsymbol{\xi}_{\alpha}\rangle
+\langle 
    \boldsymbol{\xi}_\alpha,
    \text{i}{\bf \Omega_s\times}\boldsymbol{\xi}_{\alpha}
\rangle$ and 
\begin{align}
\label{eq:Qlm}
    Q_{\ell m}^\alpha = 
    \langle \boldsymbol{\xi}_\alpha,\nabla (r^\ell Y_\ell^m)\rangle 
    % = \int_V r^\ell Y_\ell^m\rho'^*_\alpha\text{d} V
=-\frac{(2\ell+1)}{4\pi}%\frac{R_\text{eq}}{GM}
    \Phi_{\ell\alpha}
\end{align}
is the tidal overlap coefficient for mode $\alpha$.
In this work, we will consider the effect of forcing by one tidal potential $U_{\ell m}$ at a time. Stationary solutions with $\dot{c}_\alpha=-\text{i}\omega_mc_\alpha$ for each mode $\alpha$ are then
\begin{align}
\label{eq:amps}
% \notag 
%     \text{i}(\omega_{\alpha}-\omega_m)c_{\alpha}
%     &=-\frac{\text{i}e^{-\text{i}\omega_mt}}{2\epsilon_{\alpha}}
%     \sum_{n}U_{n m} Q_{n m}^\alpha 
% \\
    c_\alpha^{\ell}(t)
    &=-\frac{\text{exp}[-\text{i}\omega_mt]}
    {2\epsilon_{\alpha}(\omega_\alpha-\omega_m)}U_{\ell m}Q_{\ell m}^\alpha \\
    \notag
    &:=\hat{c}_\alpha^{\ell}(\omega_m)\text{exp}[-\text{i}\omega_mt].
\end{align}

In the quasi-adiabatic formalism \citep[e.g.,][]{Kumar1995,Burkart2012}, we can write the individual mode damping rates $\gamma_{\alpha}$ due to viscous damping as \citep[e.g.,][]{ipser1991}{}{}
$\gamma_\alpha=I_{\alpha\alpha}/(\omega_\alpha\epsilon_\alpha)$, where
\begin{equation}\label{eq:Iab}
    I_{\alpha\beta}
    % =-\frac{1}{2}\Re\int_V{\bf v}_\alpha^*\cdot[\nabla\cdot(2\mu\delta {\bf S}_\beta)]\text{d}V
    =\int_V\mu (\delta {\bf S}_\alpha^*:\delta {\bf S}_\beta)\text{d}V,
\end{equation}
where $:$ indicates contraction along both indices of rank-2 tensors. Here $\mu = \rho \nu$ is the dynamic viscosity given density $\rho$ and kinematic viscosity $\nu$, and
\begin{equation}
       \delta{\bf S}_\alpha = \frac{1}{2} \left[
        \nabla{\bf v}_\alpha
        +(\nabla{\bf v}_\alpha)^T
        -\frac{2}{3}(\nabla\cdot{\bf v}_\alpha){\bf I}
    \right]
\end{equation}
is the shear tensor for mode $\alpha$ with velocity eigenfunction ${\bf v}_\alpha=-\text{i}\omega_\alpha\boldsymbol{\xi}_\alpha$. This normalizes the viscous dissipation rate $I_{\alpha\alpha}$ by the total energy of the mode $\alpha$ \citep[][]{vick2020}. 
% We can characterize the tidal response through the effect of one tidal potential $\ell,m$ at a time, which leads to the potential Love numbers $k_{\ell m}^{n}$ such that $\delta\Phi_{\ell m}=\sum_{n}k_{\ell m}^{n}U_{nm}$, 
% \begin{equation}
%     k_{\ell m}^{n}
%     =\frac{2\pi}{(2\ell+1)}
%     \sum_{\alpha}
%     \frac{Q_{\ell m}^\alpha Q_{nm}^\alpha}
%     {\epsilon_{\alpha}(\omega_\alpha-\omega_m)}.
% \end{equation}

%dissipation integral (cite Ogilvie 2013)
%describe mode amplitudes, velocity field
In general, the time-averaged dissipation rate due to viscous dissipation of the tide is 
% eq 20 & 22 in Ipser & Lindblom come from assuming that a mode frequency has an imaginary part, whereas the tidal frequency is purely real. In general the viscous dissipation rate per unit volume is 2*mu*dS:dS (see, e.g., Bachelor's textbook), and the factor of 1/2 comes in through the time average. Ogilvie (2009) is an alternative reference, although he assumed incompressibility

\begin{equation}
\label{eq:D}
    D=\int_V\mu (\delta {\bf S}^*:\delta {\bf S})\text{d}V,
\end{equation}
where the total shear tensor $\delta {\bf S}$ is given by

\begin{align}
\label{eq:ros}
    \delta {\bf S}
    &=\frac{1}{2}\left[
        \nabla{\bf v}
        +(\nabla{\bf v})^T
        -\frac{2}{3}(\nabla\cdot{\bf v}){\bf I}
    \right].
% \\\notag
%     &=\frac{\omega_t}{2}\sum_\beta 
%     \frac{a_\beta}{\omega_\beta}
%     \left[
%         \nabla{\bf v}_\beta
%         +(\nabla{\bf v}_\beta)^T
%         -\frac{2}{3}(\nabla\cdot{\bf v}_\beta){\bf I}
%     \right]
%     \coloneqq\omega_t\sum_\beta 
%     \frac{a_\beta}{\omega_\beta} \delta{\bf S}_\beta,
\end{align}
Here the velocity fields $\bf{v}$ constitute the sum of contributions from the normal modes of stellar oscillation:
\begin{equation}
\label{eq:veldamped}
    {\bf v}
    =\sum_\alpha \hat{c}_\alpha^{\ell} {\bf v}_\alpha
    =-\sum_\alpha \frac{\omega_m}{\omega_\alpha} \frac{U_{\ell m}Q_{\ell m}^\alpha}
    {2\epsilon_{\alpha}(\omega_\alpha-\omega_m - i \gamma_{\alpha}) }{\bf v}_\alpha.
\end{equation}
% \begin{equation}
% \label{eq:vel}
%     {\bf v}
%     =\sum_\alpha c_\alpha^n {\bf v}_\alpha
%     =-\sum_\alpha \frac{U_{n m}Q_{n m}^\alpha}
%     {2\epsilon_{\alpha}(\omega_\alpha-\omega_m)}{\bf v}_\alpha
% \end{equation}90
Above, we have modified the mode amplitudes $\hat{c}_\alpha^{\ell}$ defined in Equation \ref{eq:amps} to include the individual mode damping rates $\gamma_{\alpha}$.
% , e.g.
% \begin{align}
% \label{eq:dampedamps}
% % \notag 
% %     \text{i}(\omega_{\alpha}-\omega_m)c_{\alpha}
% %     &=-\frac{\text{i}e^{-\text{i}\omega_mt}}{2\epsilon_{\alpha}}
% %     \sum_{n}U_{n m} Q_{n m}^\alpha 
% % \\
%     \hat{c}_\alpha^{\ell}
%     =-\frac{\text{exp}[-\text{i}\omega_mt]}
%     {2\epsilon_{\alpha}(\omega_\alpha-\omega_m-i\gamma_{\alpha})}U_{\ell m}Q_{\ell m}^\alpha.
% \end{align}

An important quantifier of the frequency-dependent tidal response in the perturbed body is the potential Love number $k_{\ell m}^{n}(\omega_m)$, which is a dimensionless, complex number. In this notation, for given azimuthal wavenumber $m$, the Love number describes the effect of an isolated tidal potential of harmonic degree $n$ on the gravitational response of harmonic degree $\ell$, at each tidal frequency $\omega_m$. Note that in a spherically symmetric calculation, $k_{\ell m}^{n}=0$ for $\ell \neq n$.
% in the presence of centrifugal flattening and/or the Coriolis force, k_{ell m}^n can be nonzero for ell!=n. For computing tidal torque and power, though, all that matters are the k_{ell m}^n with ell=n
%  the m=0 part does not contribute, since it produces omega_m=0, but higher-degree parts with m!=n can (see, e.g., the k31, k53, and k64 calculations in Dewberry & Lai 2022).
% In section 2.1 of that paper, n refers to different Fourier components of an eccentric orbit (rather than another spherical harmonic degree). So by "In the special case of a circular, coplanar orbit (e = i = 0), the only terms present have n = m, and l − m must be even", Gordon means that the only tidal frequencies needed are m*(Omega_o - Omega_s), rather than the more general n*Omega_o - m*Omega_s. 
The imaginary part of the Love number where $\ell=n$ is related to the energy dissipation rate $D$ by \citep{ogilvie2013}
\begin{align}
\label{eq:Dklm}
\notag
    D&=\frac{(2\ell+1)}{8\pi G}R|U_{\ell m}|^2\omega_m\mathrm{Im}[k_{\ell m}^{\ell}] \\
    &=\frac{(2\ell+1)}{8\pi}|U_{\ell m}|^2\omega_m\mathrm{Im}[k_{\ell m}^{\ell}]
\end{align}
in units of $G=M=R=1$. We make use of this relation to calculate $\mathrm{Im}[k_{\ell m}^{\ell}] $, which enters into our orbital evolution equations in \ref{sec:orbits}. Note that due to viscous coupling between modes \citep[e.g.,][]{braviner2015}{}{}, values of $\mathrm{Im}[k_{\ell m}^\ell]$ computed from \ref{eq:D} and \ref{eq:Dklm} deviate from results achieved by treating mode damping independently. We discuss this discrepancy further in Section \ref{sec:modamp}, and in a companion paper \citep{dewberry2023b}.
%double check if every equation is in the same units
\begin{figure}
    % \centering
    \includegraphics[width=\columnwidth]{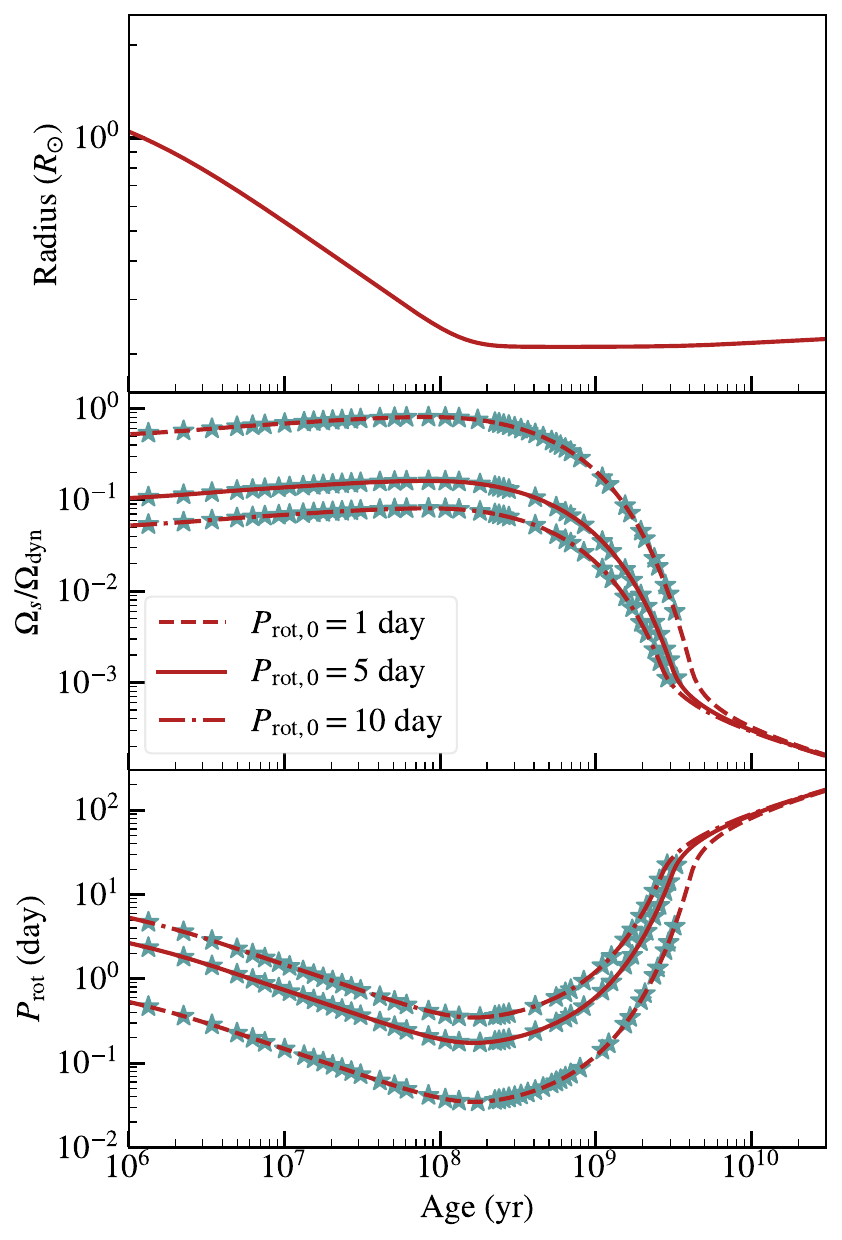}
    \caption{Evolution of the stellar radius (top panel), stellar spin frequency in units of the dynamical frequency $\Omega_{\rm dyn} =(GM/R^3)^{1/3}$ (middle panel), and stellar rotation period (bottom panel) with stellar age for the $0.2\, M_{\odot}$ stellar model used in this work. Different initial rotation periods $P_{\rm rot,0}$ of $1$, $5$, and $10$ day are shown with the linestyles indicated in the legend; the radius evolution is the same for all three initial conditions. Blue stars indicate the stellar profiles used to calculate normal modes of oscillation and construct the grid of dissipation for each $P_{\rm rot,0}$. 
    }
    \label{fig:starprops}
    
\end{figure}

\subsection{Orbital evolution due to tidal dissipation}
\label{sec:orbits}

Following \cite{ogilvie2014}, the evolution of the orbital semimajor axis $a$ and stellar spin $\Omega_s$ for circular, coplanar orbits under tidal dissipation are given by
\begin{equation}
\label{eq:aevol}
    \frac{1}{a}\frac{\mathrm{d}a}{\mathrm{d}t} = -3\, \mathrm{Im}\left[ k_{2,2}^{2}\right] \frac{M'}{M}\left(\frac{R}{a}\right)^5 \Omega_o
\end{equation}
\begin{equation}
\label{eq:spinevol}
    \frac{1}{\Omega_s}\frac{\mathrm{d}\Omega_s}{\mathrm{d}t} = \frac{3}{2}\, \mathrm{Im}\left[ k_{2,2}^{2}\right] \frac{L_o}{L_s}\frac{M'}{M}\left(\frac{R}{a}\right)^5 \Omega_o
\end{equation}
where the ratio of the orbital angular momentum to the spin angular momentum of the star is, generally,
\begin{equation}
    \frac{L_o}{L_s} = \frac{GMM'(1-e^2)^{1/2}}{I_s \Omega_s\Omega_o a}.
\end{equation}
Here $\Omega_o$ is the orbital frequency, $I_s$ is the moment of inertia of the star, and $e$ is the orbital eccentricity -- throughout this work $e=0$.

We integrate these equations to study the tidal evolution of these quantities. In the calculations described in Section \ref{sec:orbital_results}, we adopt the approach appropriate for Earth-mass planets and ignore the tidal contribution to the spin from Equation \ref{eq:spinevol}; in Section \ref{sec:reslocks}, we discuss the validity of this assumption.

The timesteps required to accurately integrate Equations \ref{eq:aevol} and \ref{eq:spinevol} are much smaller than the timescales of stellar evolution, and to recalculate the full frequency dependence of $\mathrm{Im}\left[ k_{2,2}^{2}\right]$ at each step in the orbital evolution is prohibitively expensive. As a result, for each model in a given set of snapshots of stellar evolution, we calculate the imaginary part of the Love number as a function of tidal forcing frequency, $\mathrm{Im}\left[ k_{2,2}^{2}\right](\omega_m)$. We discuss the details of this calculation and of our stellar models in Section \ref{sec:stars}. Combining the profiles of $\mathrm{Im}\left[ k_{2,2}^{2}\right](\omega_m)$ for all the stellar models provides a grid of values across tidal forcing frequency and stellar age, for a total of three dissipation grids corresponding to the initial rotation periods of $1$, $5$, and $10$ days. With each pre-calculated grid in hand, at each timestep in the orbital evolution we interpolate over the grid to find the value of $\mathrm{Im}\left[ k_{2,2}^{2}\right]$ evaluated at the tidal forcing frequency $\omega_m$ and age of that step.

\subsection{Normal Mode Oscillations of Realistic Stellar Models}
\label{sec:stars}
Using MESA \citep{mesa2011,mesa2013,mesa2015,mesa2018,mesa2019}, we simulate the evolution of a star with zero-age main sequence (ZAMS) mass $0.2\, M_{\odot}$, using three different initial rotation periods of $1$, $5$, and $10$ days respectively. We prescribe the rotational evolution of the star due to magnetic braking using the model of \cite{Matt_2015}. In this formulation of magnetic braking, the torque on the star encounters two regimes: saturated magnetic torque for small Rossby number, which applies below a critical rotation period $P_{\rm sat} \simeq 0.1 P_{\odot} \tau_{\rm cz}/\tau_{\rm cz,\odot}$; and unsaturated magnetic torque for slowly rotating stars where magnetic activity strongly correlates with the Rossby number. Here, $\tau_{\rm cz}$ is the convective turnover time. The chosen initial rotation periods of $5$ and $10$ day pass through observed values for young clusters $\sim 5$ Myr as well as older clusters at $\sim 500$ Myr and a few Gyr \citep{Matt_2015}.  %actually refer to figure

In each simulation, we evolve the star from the PMS to the terminal-age MS. For this initial mass, the star remains fully convective until $\approx 70$ Gyr, about $20$ Gyr before the terminal age MS. Figure \ref{fig:starprops} shows the evolution of the stellar radius, dimensionless stellar rotation rate, and rotation period for each of the initial conditions. As the stellar radius contracts, the star spins up until $\approx \! 100$ Myr, after which it begins to spin down to $\sim$100 day rotation periods by $10$ Gyr. The dimensionless stellar rotation rate, which is the stellar spin in units of the dynamical frequency $\Omega_{\rm dyn} =(GM/R^3)^{1/3}$, drops below $10^{-3}$ by $\approx 2.5$ Gyr. 

Throughout the PMS and MS evolution of our star, we use the 2D pseudospectral methods outlined in \cite{Dewberry2021,Dewberry2022,Dewberry2023} to compute normal modes of stellar oscillation in order to calculate the tidal response. %reference appropriate equations and briefly describe SPORT
These non-perturbative methods for finding stellar oscillation modes, which can capture the full effects of the Coriolis force by treating the linearized fluid equations as non-separable in ($r,\theta$) under rotation, assume rigid rotation and zero viscosity. The axisymmetry of the background state allows modes to be labeled by unique azimuthal wavenumbers $m$.
Assuming spherical symmetry, the 1D, radial profiles of realistic stellar structure from MESA constitute the equilibrium profiles for the 2D pseudospectral methods, which we use to solve for adiabatic normal modes with time dependence $\exp{[-i\omega_{\alpha} t]}$ for rotating frame mode frequencies $\omega_{\alpha}$.
%could say that we project the 1d MESA profiles onto a 2d grid in radius and \mu = cos \theta

% \begin{figure}
%     \centering
%     \includegraphics[width=\columnwidth]{Prot1_Qlm_freqs.pdf}

%     \caption{Evolution of the inertial-frame mode frequencies $\sigma$ in units of the dynamical frequency (top panel) and the absolute value of the $\ell=m=2$ tidal overlap coefficients $Q_{\ell,m}$ (bottom panel) for a subset of f-modes and i-modes during the pre-MS to MS evolution of the $P_{\rm rot,0} = 1$ day stellar model. 
%     }
%     \label{fig:QlmSigmasHi}
    
% \end{figure}

In this work, we solve for modes with azimuthal wavenumber $m=2$ and thus consider tidal forcing frequencies $\omega_m = 2(\Omega_o-\Omega_s)$. We keep track of the $\ell=2,4,6$ prograde and retrograde f-modes, where $\ell$ is the spherical harmonic degree, as well as 12 i-modes. Here, prograde refers to modes whose inertial-frame frequency $\omega_{\alpha}$ is positive, and retrograde modes have  $\omega_{\alpha} <0$. For the i-modes, we characterize them by the number of nodes in directions that are roughly horizontal ($n_1$) and vertical ($n_2$), where horizontal describes the direction of the cylindrical radius $R$, and vertical is the cylindrical $z$ direction \citep{wu2005a}.
We consider the $n_1+n_2=1$, $n_1+n_2=2$, and $n_1+n_2=3$ prograde and retrograde i-modes. To refer to the modes in the text, we follow the notation of \cite{Dewberry2022} so that $f_{\ell,m,\pm}$ denotes the f-modes and $i_{m,n_1,n_2,\pm}$ labels the i-modes, with $+$ and $-$ indicating prograde or retrograde modes respectively. Inertial modes are only labeled prograde or retrograde if the $m, n_1,n_2$ values are duplicated in another mode. The chosen set of i-modes are the longest-wavelength modes; shorter-wavelength modes are unlikely to contribute significantly to tidal dissipation as they have smaller tidal overlap coefficients $Q^\alpha_{\ell m}$ and smaller amplitudes.

For each stellar profile, we calculate the dissipation $D$ from Equation \ref{eq:D}, with Equations \ref{eq:ros} and \ref{eq:veldamped} summed over the oscillation modes $\alpha$ computed for each stellar model. In each sum, we include the f-modes and i-modes listed above. Equation \ref{eq:D} involves a double sum over the mode velocity eigenfunctions that allows us to capture the dissipative coupling between different modes, e.g. $I_{\alpha\beta}$ (Equation \ref{eq:Iab}). In most prior work that takes the modal decomposition approach, they instead perform a single sum over each mode to calculate tidal dissipation (using, e.g., Equation \ref{eq:klmsum}) that does not incorporate this coupling. However, we find that the terms $I_{\alpha,\beta}$ can be quite large and contribute significantly to the dissipation away from mode resonances. Given $D$, the imaginary part of the Love number $\mathrm{Im}\left[ k_{2,2}^{2}\right]$ as a function of tidal forcing frequency $\omega_m$ can be found from Equation \ref{eq:Dklm}.

\begin{figure}
    % \centering
    \includegraphics[width=\columnwidth]{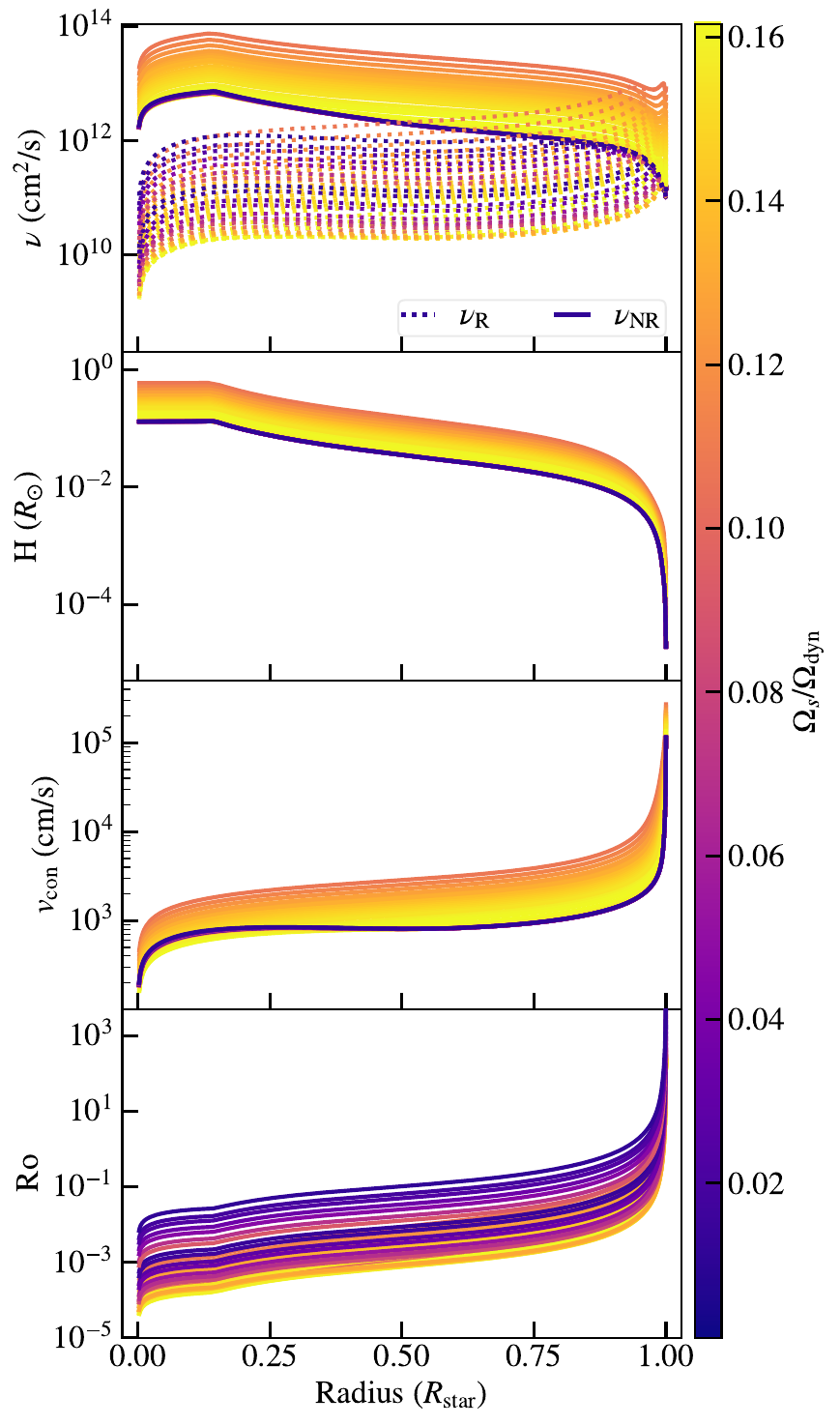}
    \caption{ The kinematic viscosity (first panel), scale height (second panel), convective velocity (third panel), and Rossby number (fourth panel) versus radial coordinate (scaled by the total stellar radius) for a set of profiles during the evolution of a $P_{\rm rot,0}=5$ day model. The top panel shows the turbulent convective viscosity with ($\nu_{\rm R}$, Equation \ref{eq:Rvisc}) and without ($\nu_{\rm NR}$, Equation \ref{eq:NRvisc}) a rotational correction. As Equation \ref{eq:Rvisc} is a function of Rossby number, $\nu_{\mathrm{R}}$ varies with rotation and approaches the value of $\nu_{\mathrm{NR}}$ towards the surface of the star where the Rossby number is large. On the main sequence, $\nu_{\mathrm{NR}}$, $H$, and $v_{\rm con}$ remain nearly constant so all lines lie on top of each other under the purple line. 
    }
    \label{fig:viscprops}
    
\end{figure}

\subsection{Viscous dissipation in convective zones}
To treat the viscous dissipation of the tides, we estimate the effective turbulent viscosity of convection throughout the fully convective models. The kinematic eddy viscosity $\nu$ of turbulent convective friction can be described using mixing length theory as \citep{zahn1966b}:
%describe two ways of estimates
\begin{equation}
\label{eq:NRvisc}
    \nu_{\mathrm{NR}} \sim v_{\rm con} l_{\rm con} \sim \left(\frac{L_{\rm con}}{4\pi\rho r^2}\right)^{1/3} H,
\end{equation}
where $v_{\rm con}$, $l_{\rm con}$, and $L_{\rm con}$ are the convective velocity, mixing length, and convective luminosity respectively, $H$ is the pressure scale height, and $\rho$ is the density.
This simple estimate ignores the effect of rotation on convective flows, as well as the attenuation of the dissipation efficiency when the tidal forcing frequency is much larger than the convective turnover frequency, $\omega_m \gg \omega_{\rm con} \sim v_{\rm con}/l_{\rm con}$ (see, e.g., \citealt{zahn1966b,goldreich1977,duguid2020a}).
%Stronger turbulent damping has been proposed \citep{terquem2021}, but we agree with the conclusion of \cite{barker2021} that the relevant terms vanish upon integration over the volume of the star.

Figure \ref{fig:viscprops} shows the variation of $l_{\rm con}=H$, $v_{\rm con}$, and the kinematic eddy viscosity $\nu_{\mathrm{NR}}$ (solid lines in top panel) with radial coordinate, colored by the dimensionless stellar rotation rate for the $P_{\rm rot,0}=5$ day models. While the star is on the pre-MS, the scale height and convective velocity are elevated because the stellar radius and luminosity are much larger than on the MS, leading to higher viscosity on the pre-MS (solid yellow and orange curves in Figure \ref{fig:viscprops}). While on the MS, the stellar radius remains nearly constant, so the scale height, convective velocity, and thus $\nu_{\mathrm{NR}}$ all remain similar throughout the rest of the star's lifetime as it spins down to small values of $\Omega_s/\Omega_{\rm dyn}$ (solid purple curves).

However, rapid rotation such as that experienced by our stellar models affects turbulent convective flows, in particular by inhibiting the efficiency of the turbulent friction. \cite{mathis2016} address how the viscosity in Equation \ref{eq:NRvisc} should be modified for the rotating case by scaling the convective velocity and mixing length for rapid rotation to the non-rotating case \citep{stevenson1979,barker2014}:
\begin{align}
    \frac{v_{\rm con}(\mathrm{Ro})}{v_{\rm con}(\Omega_s=0)} &= 1.5(\mathrm{Ro})^{1/5} \\
    \frac{l_{\rm con}(\mathrm{Ro})}{l_{\rm con}(\Omega_s=0)} &= 2(\mathrm{Ro})^{3/5}.
\end{align}
Here $\mathrm{Ro} = \omega_{\rm con}/\Omega_s$ is the Rossby number. \cite{mathis2016} use the above scalings for the rapidly rotating regime of $\mathrm{Ro} \lesssim 0.25$, whereas in the slowly rotating regime of $\mathrm{Ro} \gtrsim 0.25$ they employ the following:
\begin{align}
    \frac{v_{\rm con}(\mathrm{Ro})}{v_{\rm con}(\Omega_s=0)} &= 1-\frac{1}{242(\mathrm{Ro})^{2}} \\
    \frac{l_{\rm con}(\mathrm{Ro})}{l_{\rm con}(\Omega_s=0)} &= \left(1+\frac{1}{82(\mathrm{Ro})^{2}}\right)^{-1}.
\end{align}
With these scaling relations, the kinematic eddy viscosity in the rotating case may be estimated as 
\begin{align}
\label{eq:Rvisc}
    \nu_{\mathrm{R}}(\mathrm{Ro}) &\sim v_{\rm con}(\mathrm{Ro}) l_{\rm con}(\mathrm{Ro}) \\
    \notag
    &\sim \nu_{\mathrm{NR}} \frac{v_{\rm con}(\mathrm{Ro})}{v_{\rm con}(\Omega_s=0)}\frac{l_{\rm con}(\mathrm{Ro})}{l_{\rm con}(\Omega_s=0)}. 
\end{align}

As expected from the rotational evolution of the star (Figure \ref{fig:starprops}), $\mathrm{Ro}$ is much smaller throughout the star during the pre-MS than the late MS.
From the bottom panel of Figure \ref{fig:viscprops}, we see that $\mathrm{Ro}\ll 1$ in the stellar interior, but increases sharply towards the surface as $H$ drops and $v_{\rm con}$ rises. As a result, the viscosity transitions between the rapidly-rotating and slowly-rotating regimes within the star. 
The dotted lines in the top panel of Figure \ref{fig:viscprops} illustrate  that the effect of rapid rotation in our models is to lower the kinematic viscosity in the majority of the star, as $\mathrm{Ro}$ is small except towards the surface. In addition, the rotationally-corrected viscosity now varies on the MS due to the dependence on $\mathrm{Ro}$, so that the slowly-rotating models towards the late MS $\gtrsim 1$ Gyr have 1--2 orders of magnitude larger $\nu_{\mathrm{R}}$ than at peak rotation around $\sim 10$--$100$ Myr.

\begin{figure*}
    \centering
    \includegraphics[width=0.495\textwidth]{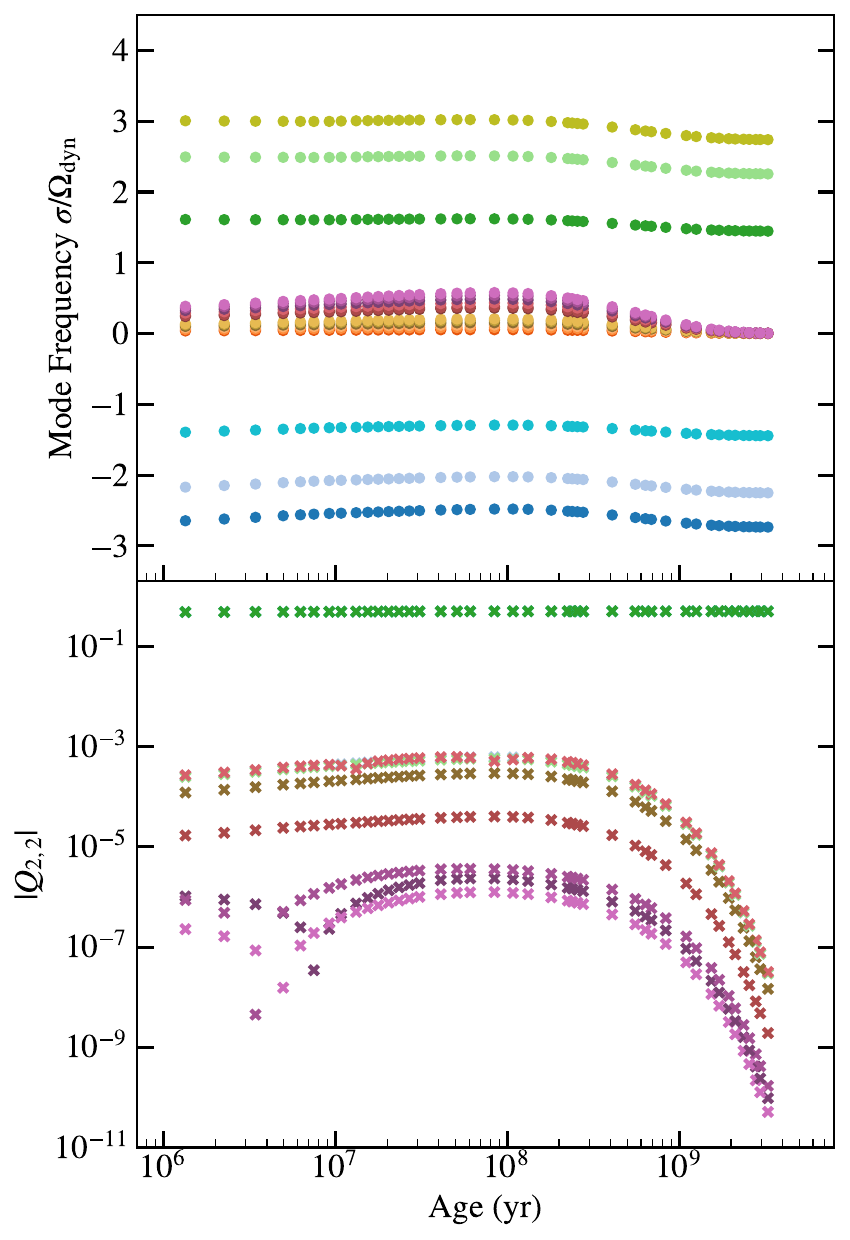}
     \includegraphics[width=0.495\textwidth]{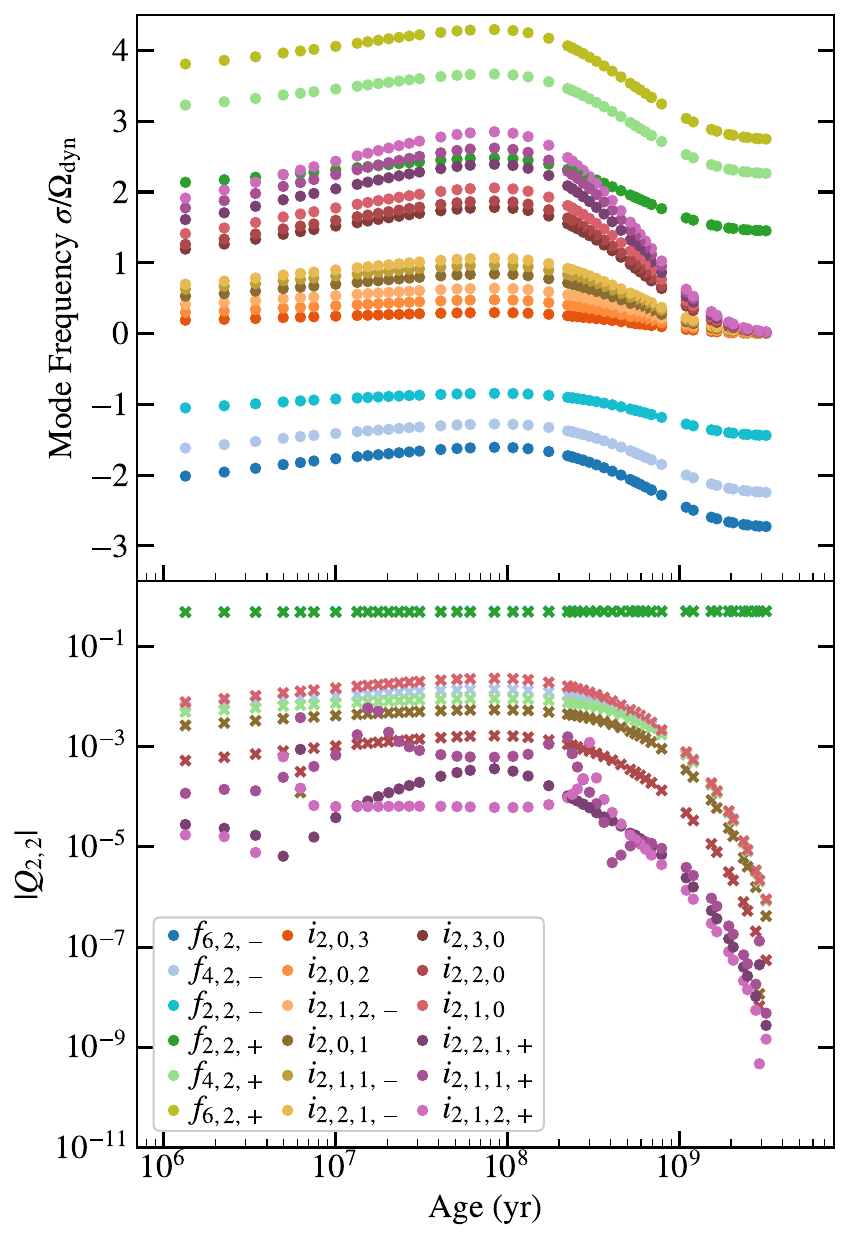}

    \caption{Evolution of the inertial-frame mode frequencies $\sigma$ in units of the dynamical frequency  $\Omega_{\rm dyn}$ (top panel) and the absolute value of the $\ell=m=2$ tidal overlap coefficients $Q_{\ell,m}$ (bottom panel) for the set of f-modes and i-modes we include during the pre-MS to MS evolution of the $P_{\rm rot,0} = 5$ day (left) and $1$ day (right) sets of stellar models. 
    In the bottom panel, a subset of these modes are plotted. For the $P_{\rm rot,0} = 1$ day models, i-modes that experience avoided crossings are plotted with a circle marker in the bottom panel.
    }
    \label{fig:QlmSigmas}
    
\end{figure*}

A second factor not implemented here is the frequency-dependent attenuation of the efficiency for rapid tide where $\omega_m \gg \omega_{\rm con}$. In the literature, several scalings have been proposed for this reduction, including linear \citep{zahn1966b,penev2007,penev2009} or quadratic \citep{goldreich1977,goodman1997,ogilvie2012,duguid2020a}, with some arguing that no reduction is appropriate \citep{terquem2023}. For i-modes, whose frequencies scale with the stellar rotation rate, the rapid tide attenuation will effectively scale as $\mathrm{Ro}$ or $\mathrm{Ro^2}$ depending on which prescription is taken. In the former case, we expect similar effects on the mode damping compared to the considerations of rotating convection described above, but in the case of quadratic scaling, the suppression will be more dramatic than shown here. Furthermore, if both reductions act simultaneously, the viscosity would strongly scale with $\mathrm{Ro}$ and decrease even more. 

We repeat our calculation of the dissipation as described in Section \ref{sec:stars} for both assumptions of the kinematic viscosity $\nu_{\rm R}$ or $\nu_{\rm NR}$ as described above. The fiducial results shown throughout this work use the results from assuming $\nu_{\rm R}$, and we address comparisons between the two assumptions in Sections \ref{sec:viscdiss} and \ref{sec:orbvisc}.

\section{Tidal dissipation across stellar evolution}
\label{sec:results}

% \begin{figure*}
%     % \centering
%     \includegraphics[width=\textwidth]{k22_all_sideways_some.png}
%     % \includegraphics[width=\textwidth]{k22_all.png}
%     \caption{The absolute value of $\mathrm{Im}\left[k_{2,2}^2\right]$ as a function of tidal forcing frequency (in dynamical units) and stellar age, as calculated from Equations \ref{eq:D}--\ref{eq:Dklm} using the f-modes and i-modes listed in the legend. Each panel shows the result for a different $P_{\rm rot,0}$. The dotted lines overplotted show the evolution of the rotating-frame mode frequencies $\omega_{\alpha}$, which line up with maxima of the dissipation $\mathrm{Im}\left[k_{2,2}^2\right]$ throughout the star's lifetime due to resonances.
%     }
%     \label{fig:k22all}
    
% \end{figure*}

\begin{figure*}
    % \centering
    \includegraphics[width=\textwidth]{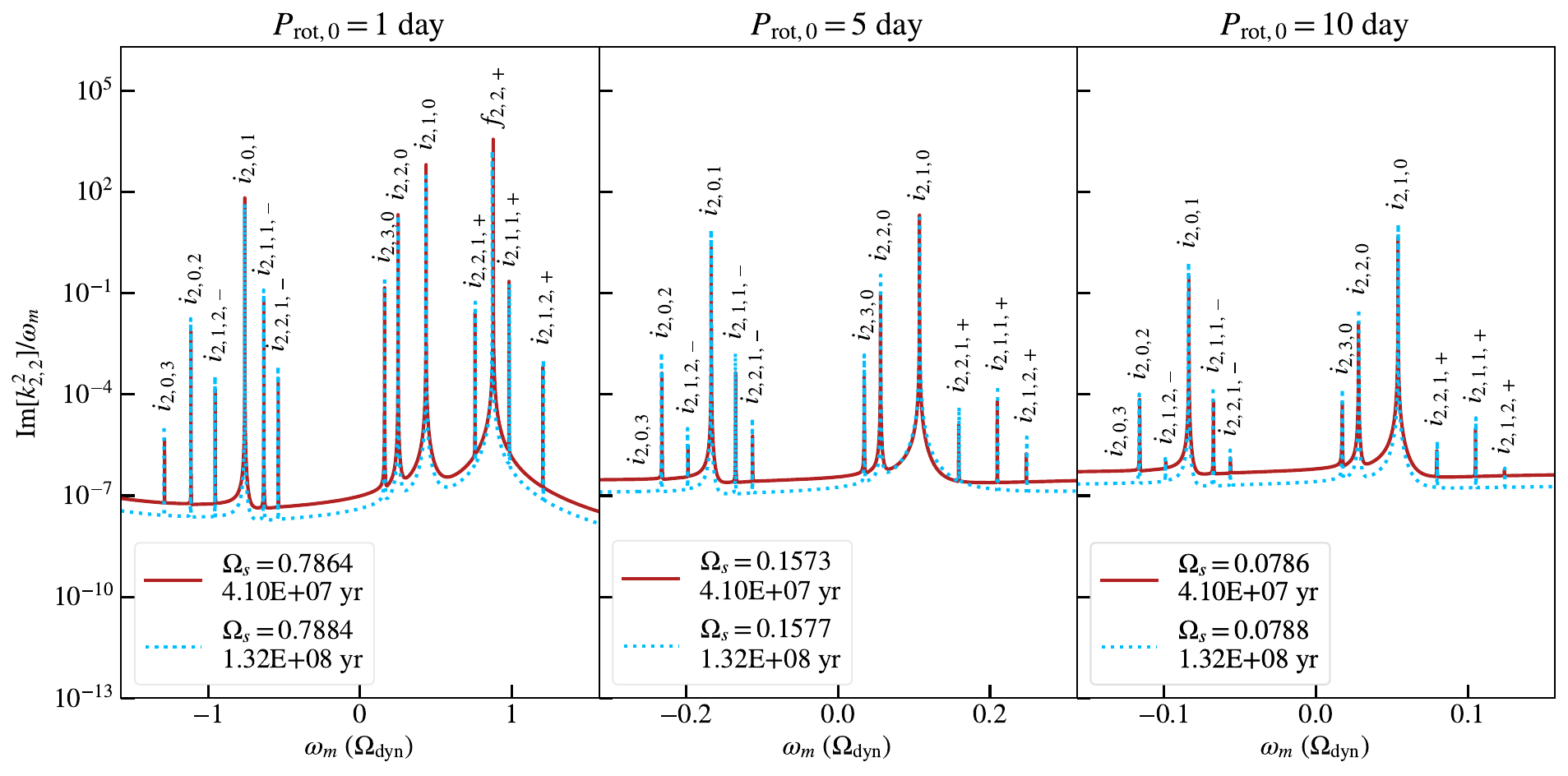}
    \caption{Imaginary parts of $k_{2,2}^2$ as a function of tidal forcing frequency $\omega_m$ (in units of the dynamical frequency $\Omega_{\rm dyn}$). The y-axis is scaled by $\omega_m$, and panels are labeled by the value of $P_{\rm rot, 0}$ used to evolve the stellar model. In each panel, the red lines show the dissipation on the pre-MS for the stellar rotation rates $\Omega_s$ (in units of  $\Omega_{\rm dyn}$) listed in the legend, whereas the blue dotted lines show the same for a model on the MS with a very similar rotation rate. The x-axis limits in each panel are limited to $[-2\Omega_s,2\Omega_s]$ to show the inertial mode range. Each resonant peak is labeled by the mode responsible for the resonance. The dissipation is larger on the pre-main sequence due to larger turbulent convective viscosity.
    }
    \label{fig:k22omm1d}
    
\end{figure*}

\subsection{Mode Evolution with Stellar Evolution}
\label{sec:dissipation_results}
%emphasize evolution of mode spectrum over time, pointing out avoided crossings that occur
%show Qlm evolution for test_new_methods to show avoided crossing. but this may not affect the orbital evolution
The top panel of Figure \ref{fig:QlmSigmas} shows the evolution of the inertial-frame mode frequency $\sigma$ for the set of f-modes and i-modes tracked in this work, for initial conditions $P_{\rm rot,0}=5$ and $1$ day. The description for $P_{\rm rot,0} = 5$ day is representative of $P_{\rm rot,0}=10$ day as both are calculated for slowly-rotating stars. As the stellar spin increases and decreases (see Figure \ref{fig:starprops}), the i-mode frequencies are generally proportional to the stellar spin and follow the same evolution. For the slower-spinning models, the f-mode frequencies also increase and decrease with stellar spin slightly, but remain fairly constant and well separated from the i-mode frequencies. In the $P_{\rm rot,0}=1$ day model on the right, the i-mode frequencies again increase and decrease with rotation, but here the f-mode frequencies also show a significant increase with rotation rate. 
As the star spins down to $\Omega_s/\Omega_{\rm dyn} \sim 10^{-3}$ at $\sim 2.5$ Gyr, the i-mode frequencies similarly approach $0$ in both models. 

The bottom panels show the absolute value of the $\ell=m=2$ tidal overlap coefficients $|Q_{2,2}|$ (Equation \ref{eq:Qlm}), which affect the mode amplitudes; for all i-modes, these drop by several orders of magnitude once the star spins down to $\Omega_s/\Omega_{\rm dyn}  \sim 10^{-3}$ at a few Gyr. Throughout the evolution, the prograde and retrograde $\ell=2$ f-modes still have much larger $|Q_{2,2}|$, but the next largest values come from the longest-wavelength i modes, e.g. $i_{2,0,1}$ and $i_{2,1,0}$, as well as the $\ell=4$, $m=2$ f-modes.

In the $P_{\rm rot,0}=1$ day model, the prograde $\ell=2,m=2$ f-mode frequency crosses those of the highest-frequency i-modes at large rotation rates $\Omega_s \approx 0.6$--$0.8\, \Omega_{\rm dyn}$. %look up the specific rotation rate of the avoided crossing
This leads to several instances of an ``avoided crossing", during which the f-mode and i-mode mix in character as they approach one another in frequency \citep[see][for a discussion of similar i-mode/f-mode mixing in isentropic polytropes]{Dewberry2022}. 
These avoided crossings occur between the f-mode $f_{2,2,+}$ and the i-modes $i_{2,1,1,+}$ and $i_{2,1,2,+}$, which are distinguished with dots instead of crosses in the bottom right panel of Figure \ref{fig:QlmSigmas}. The i-mode $i_{2,2,1,+}$ also approaches $f_{2,2,+}$ in frequency near the maximum stellar rotation rate (also shown as dots). These three i-modes show increases in $|Q_{2,2}|$ over several orders of magnitude as their mode frequencies approach or cross that of the f-mode. For the $i_{2,2,1,+}$ i-mode whose frequency nears, but does not cross, that of the f-mode, the enhancement is reduced yet still visible. Though none of these i-modes achieve $|Q_{2,2}|$ values that are comparable to the f-mode's tidal overlap integral, the effect of mode mixing allows these relatively short wavelength i-modes to contribute similar $|Q_{2,2}|$ coefficients compared to the longest-wavelength i-modes, whereas they would otherwise have negligibly small $|Q_{2,2}|$ in comparison. The consequence, as we discuss below in the context of Figure \ref{fig:k22omm1d}, is to enhance the amplitudes of these modes during these periods of rapid rotation, leading to greater resonant dissipation. 
%show dissipation functions w/spikes

\begin{figure}
    \centering
    \includegraphics[width=\columnwidth]{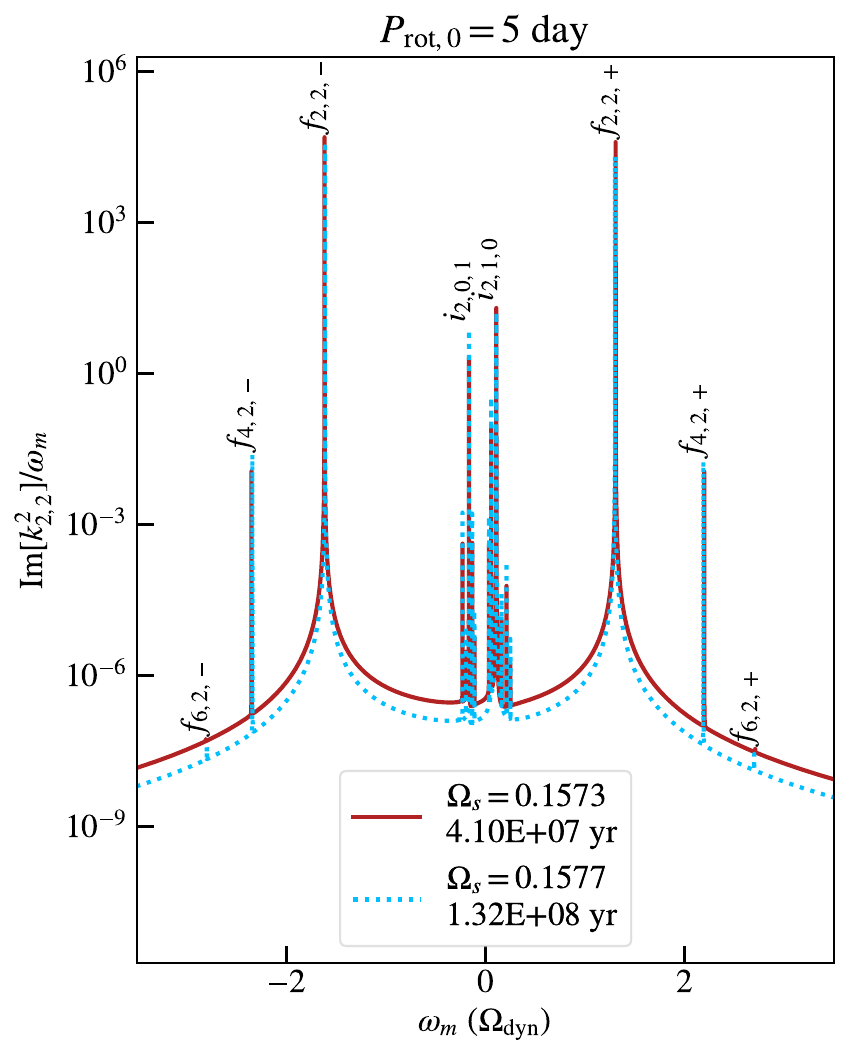}
    \caption{Same as Figure \ref{fig:k22omm1d}, but zoomed out to show the f-modes for the $P_{\rm rot,0}=5$ day models. The f-mode and largest i-mode resonances are labeled by the mode responsible for each resonance. }
    \label{fig:k22omm1d_zoomedout}
    
\end{figure}

\subsection{Tidal dissipation coupled with stellar evolution}
%discuss smooth plot
%Given the mode spectrum over time in each set of stellar models, 
We compute the frequency-dependent dissipation $D$ from Equation \ref{eq:D} in each stellar model, using the f-modes and i-modes shown in Figure \ref{fig:QlmSigmas} to construct the velocity fields of Equation \ref{eq:veldamped}. Models shown in this section use the rotation correction to viscosity of Equation \ref{eq:Rvisc} unless specified otherwise. 
%By inverting Equation \ref{eq:Dklm}, we find $\mathrm{Im}\left[k_{2,2}^2\right]$ across forcing frequency and age. 
% An overview of how the magnitude of $\mathrm{Im}\left[k_{2,2}^2\right]$ evolves over time is shown in Figure \ref{fig:k22all}. Darker contours in Figure \ref{fig:k22all} represent the trajectory of different mode resonances that cause amplified dissipation at tidal forcing frequencies $\omega_m$ that equal the rotating-frame frequency $\omega_{\alpha}$ of a stellar oscillation mode. This enhancement can be understood from the difference $\omega_m - \omega_{\alpha}$ that appears in the denominator of the velocity amplitudes (\ref{eq:veldamped}).
% As a result, the paths of $\omega_{\alpha}$ for the normal modes plotted in Figure \ref{fig:k22all} align with curves of heightened dissipation. 

%discuss spiky plot
Figures \ref{fig:k22omm1d}--\ref{fig:k22omm1d_zoomedout} show the positive-definite quantity $\mathrm{Im}\left[k_{2,2}^2\right]/\omega_m$ versus $\omega_m$ at different snapshots in the stellar evolution. In each panel corresponding to different $P_{\rm rot,0}$, the spikes that rise orders of magnitude above the baseline dissipation are peaks of resonance with the stellar oscillation mode as labeled. Modes with shorter peaks are shorter-wavelength and have smaller maximum amplitudes due to damping (Equation \ref{eq:veldamped}) and smaller tidal overlap integrals (Equation \ref{eq:Qlm}).
Note that $\mathrm{Im}\left[k_{2,2}^2\right]/\omega_m$ is constant across $\omega_m=0$ because of the scaling by $\omega_m$. The value of $\mathrm{Im}\left[k_{2,2}^2\right]$ actually
approaches zero at $\omega_m = 0$, which corresponds to the case of corotation ($\Omega_s = \Omega_o$).  

In red, the stellar model shown is on the pre-MS, whereas the blue dashed curve is for a model on the MS with a similar rotation rate (in units of $\Omega_{\rm dyn}$). Compared to the pre-MS model, the more compact star on the MS has a lower viscosity and therefore smaller off-resonant dissipation even for the same dimensionless rotation rate. For the models shown here, the difference amounts to a factor of a few, but similar comparisons involving the more extended star on the early pre-MS can differ by up to an order of magnitude. 

Figure \ref{fig:k22omm1d} focuses on the region $-2\Omega_s < \omega_m < 2\Omega_s$, which is the range over which inertial modes are excited. 
For $P_{\rm rot,0}=1$ day at the rotation rate of $\Omega_s/\Omega_{\rm dyn} \sim 0.8$ shown, the peak for $f_{2,2,+}$ falls within this range and is very near in frequency to $i_{2,2,1+}$ and $i_{2,1,1,+}$. In the vicinity of this rotation rate, the avoided crossings between the f-mode and i-modes as described in the context of Figure \ref{fig:QlmSigmas} amplify the i-mode dissipation. In contrast, the same two i-modes in slower-rotating models appear with some of the smallest resonant spikes, as they have quite short wavelengths and small $|Q_{2,2}|$ values. Throughout the evolution, all the modes we consider for  $P_{\rm rot,0}=1$ day augment the dissipative response significantly near resonance, but fewer modes contribute significantly for the slower-rotating stars. This corresponds to overall smaller magnitudes of $|Q_{2,2}|$ exhibited by the slower-rotating models.

Figure \ref{fig:k22omm1d_zoomedout} shows a larger range in tidal forcing frequency for the $P_{\rm rot,0}=5$ day model. In this region, we can see that the f-mode amplitudes dominates the dissipation away from resonant peaks, which can also be thought of as the equilibrium tidal response.  As the initial rotation period increases, the dissipation away from resonance mildly increases due to the rotational correction to viscosity. 
% The $\ell=2$ f-modes are by far the largest peaks compared to any other mode, with the next largest contributions from the longest-wavelength i-modes; the $\ell = 4$ f-modes are much smaller, and the $\ell=6$ f-modes contribute negligibly. 
%As the models with $P_{\rm rot,0}=1$ day have a few--10 times larger dimensionless rotation rates, the i-mode resonances span a larger range in tidal forcing frequency and also have taller peaks, with $i_{2,2,0}$ in particular joining $i_{2,1,0}$ and $i_{2,0,1}$ in rising above the $\ell=4$ f-modes. 
The largest dissipation comes from $\ell=m=2$ f-modes and the longest-wavelength i-modes ($i_{2,1,0}$ and $i_{2,0,1}$). In older models with slower rotation rates, the strength of resonant i-mode dissipation fades as the star spins down greatly on the MS, tracking the rapid decline of $Q_{2,2}$ with age as seen in Figure \ref{fig:QlmSigmas}. The $\ell=m=2$ f-mode peaks maintain similar heights throughout the evolution.

As the star spins up from a few--$100$ Myr, the viscosity grows more inefficient due to the decreasing Rossby number, and the dissipation away from resonance correspondingly decreases. Once the star spins down significantly, the Rossby number and consequently the viscosity grow. 
The non-resonant dissipation is similar on the pre-MS at $\lesssim 10$ Myr and on the late MS at $\sim 2$ Gyr, even though the star is spinning 10--100 times faster on the pre-MS. Since the fully convective star has a much larger radius on the pre-MS, the convective velocity and scale height are correspondingly larger, leading to larger viscosities on the pre-MS for a given rotation rate.

\begin{figure}
    \centering
    \includegraphics[width=\columnwidth]{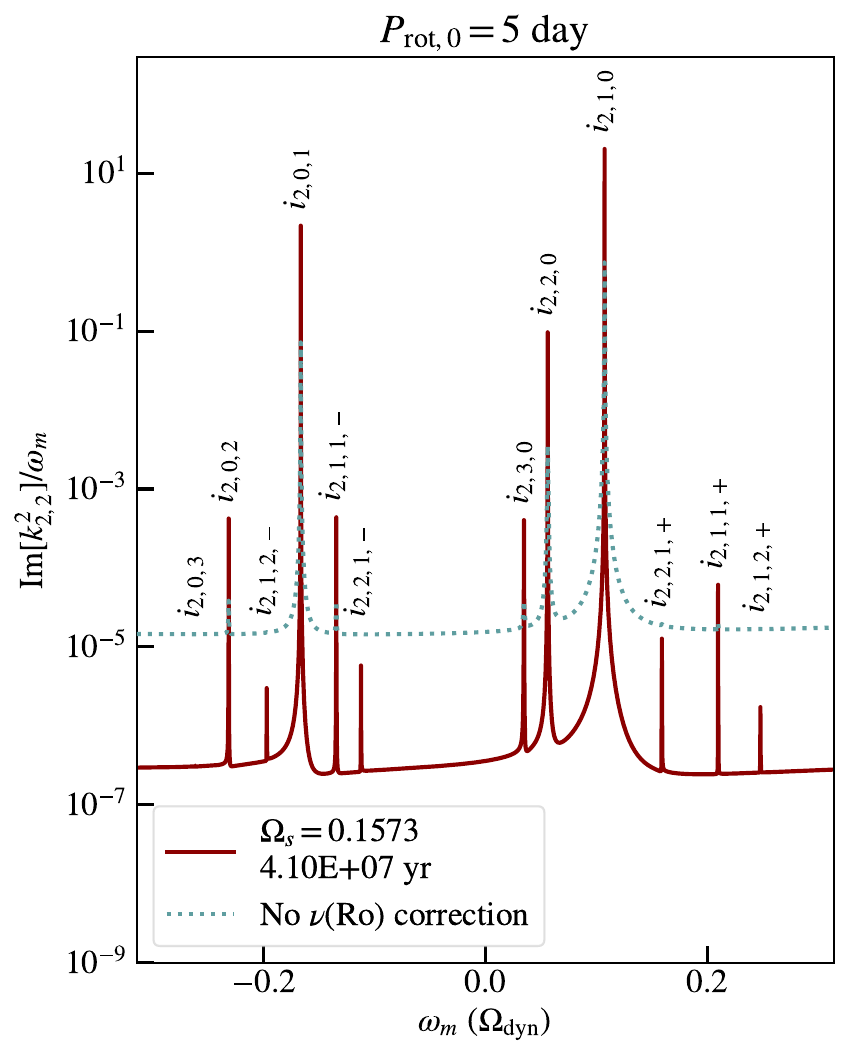}
    \caption{Similar to the center panel of Figure \ref{fig:k22omm1d} ($P_{\rm rot,0}=5$ day), $\mathrm{Im}[k_{2,2}^2]$ scaled by $\omega_m$ as a function of tidal forcing frequency  $\omega_m$ is shown, but now to compare the effect of viscosity. The red line shows the dissipation using the correction to viscosity due to rotation (Equation \ref{eq:Rvisc}). The blue dotted lines show the same but using viscosity without the rotation correction (Equation \ref{eq:NRvisc}). 
    Including the effects of rotation lowers the viscosity, leading to sharper resonant peaks, but smaller off-resonant dissipation by $1$--$2$ orders of magnitude.
    }
    \label{fig:k22comparevisc}
    
\end{figure}

\subsubsection{Comparison of different assumptions for the viscosity}
\label{sec:viscdiss}

We compare in Figure \ref{fig:k22comparevisc} the effects of including or excluding the correction to viscosity under rapid rotation. In red, the quantity $\mathrm{Im}\left[k_{2,2}^2\right]/\omega_m$ is shown for the same pre-MS model as in Figures \ref{fig:k22omm1d} and \ref{fig:k22omm1d_zoomedout}, for which we compute the dissipation assuming the kinematic viscosity to be $\nu_{\mathrm R}$ (Equation \ref{eq:Rvisc}). In blue, the same model is shown for a calculation using $\nu_{\mathrm{NR}}$ (Equation \ref{eq:NRvisc}) instead. As seen in Figure \ref{fig:viscprops}, $\nu_{\mathrm{NR}} \gg \nu_{\mathrm{R}}$ throughout the majority of the star, and this impacts the dissipation curves in two significant ways. The off-resonance dissipation is orders of magnitude larger when using the $\nu_{\mathrm{NR}}$ instead of $\nu_{\mathrm{R}}$, which follows from the attenuation of $\nu_{\mathrm{R}}$ at small $\mathrm{Ro}$ by $\approx 2$--$3$ orders of magnitude. Furthermore, the resonant spikes are wider and shorter for the higher viscosity, $\nu_{\mathrm{NR}}$. As discussed in Section \ref{sec:orbvisc}, we expect larger off-resonance dissipation to cause more rapid orbital migration as a result of the equilibrium tide, whereas smaller peak heights decrease the likelihood of resonance locking for orbital evolutionary calculations that utilize the $\mathrm{Ro}$-independent viscosity. We note that if the additional effect of frequency-dependent attenuation of the viscous efficiency were to be included such that the viscosity scaled even more strongly with $\mathrm{Ro}$, then the profiles of $\mathrm{Im}\left[k_{2,2}^2\right]/\omega_m$ would trend towards even smaller dissipation away from resonance and narrower, taller peaks.
%plot dissipation functions with spikes comparing different viscosities
%mention that with the frequency dependent attenuation, could see even larger spikes, etc.

\begin{figure*}

    \includegraphics[width=\textwidth]{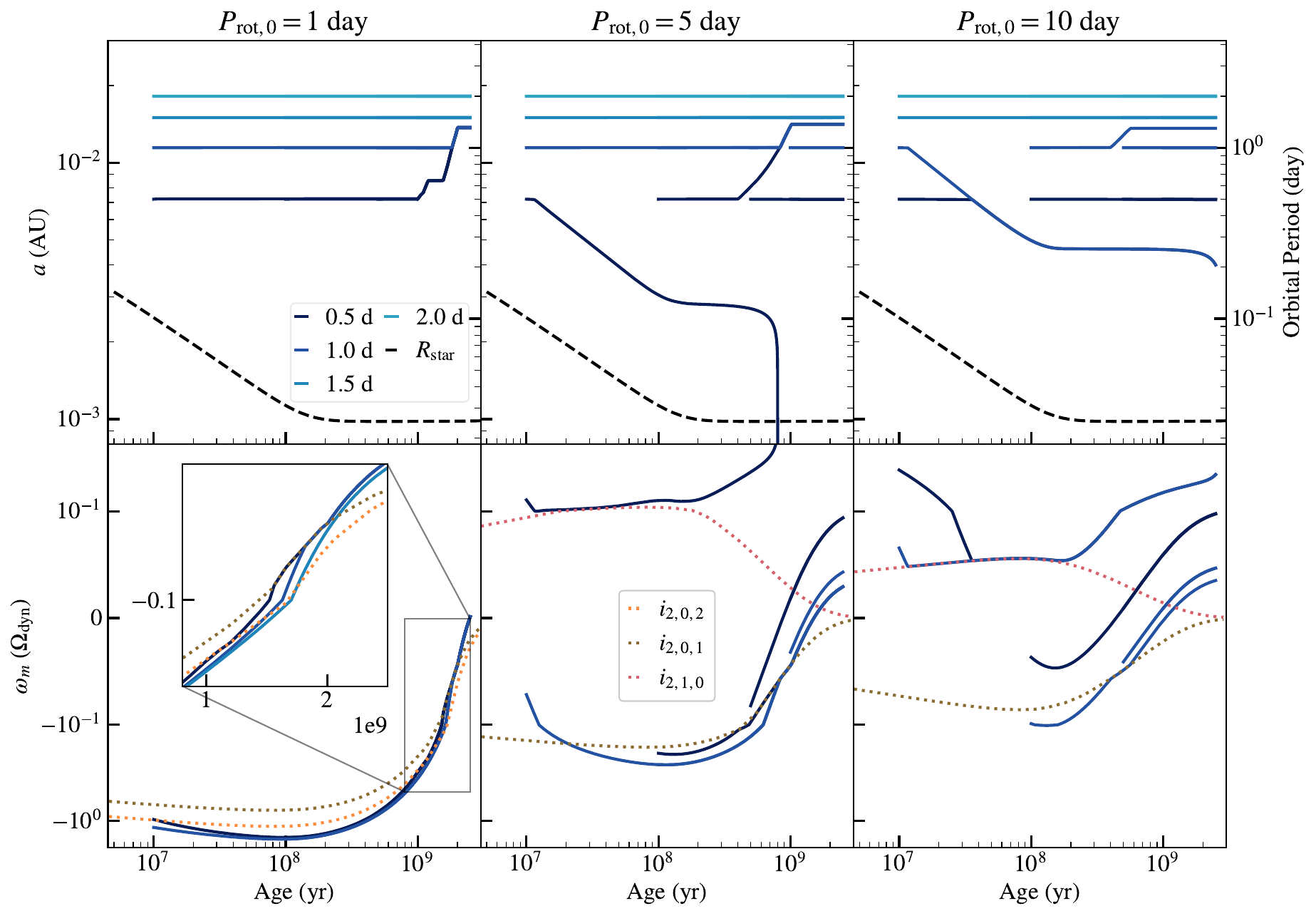}
    \caption{\textbf{Top row}: Evolution of the semimajor axis or orbital period of a $M_p = M_e$ Earth-mass companion for the different initial orbital periods shown in the legend in the upper left. Each column depicts results from each stellar evolutionary model with initial rotation periods $P_{\rm rot,0}$ as labeled, and the dissipation is calculated using the correction to viscosity under rotation (Equation \ref{eq:Rvisc}). The dashed black line shows the radius of the star.
    \textbf{Bottom row}: The tidal forcing frequency $\omega_m$ (in units of the dynamical frequency) for the same models.
    Dotted lines show the rotating-frame mode frequencies $\omega_{\alpha}$ for the i-modes listed in the legend in the center panel. Significant migration occurs when the companion's tidal forcing frequency enters a resonance lock with a stellar oscillation mode, which is visualized as a solid line overlapping with a dotted line in the bottom panels.
    }
    \label{fig:Meorbitsummary}
    
\end{figure*}

\begin{figure*}

    \includegraphics[width=\textwidth]{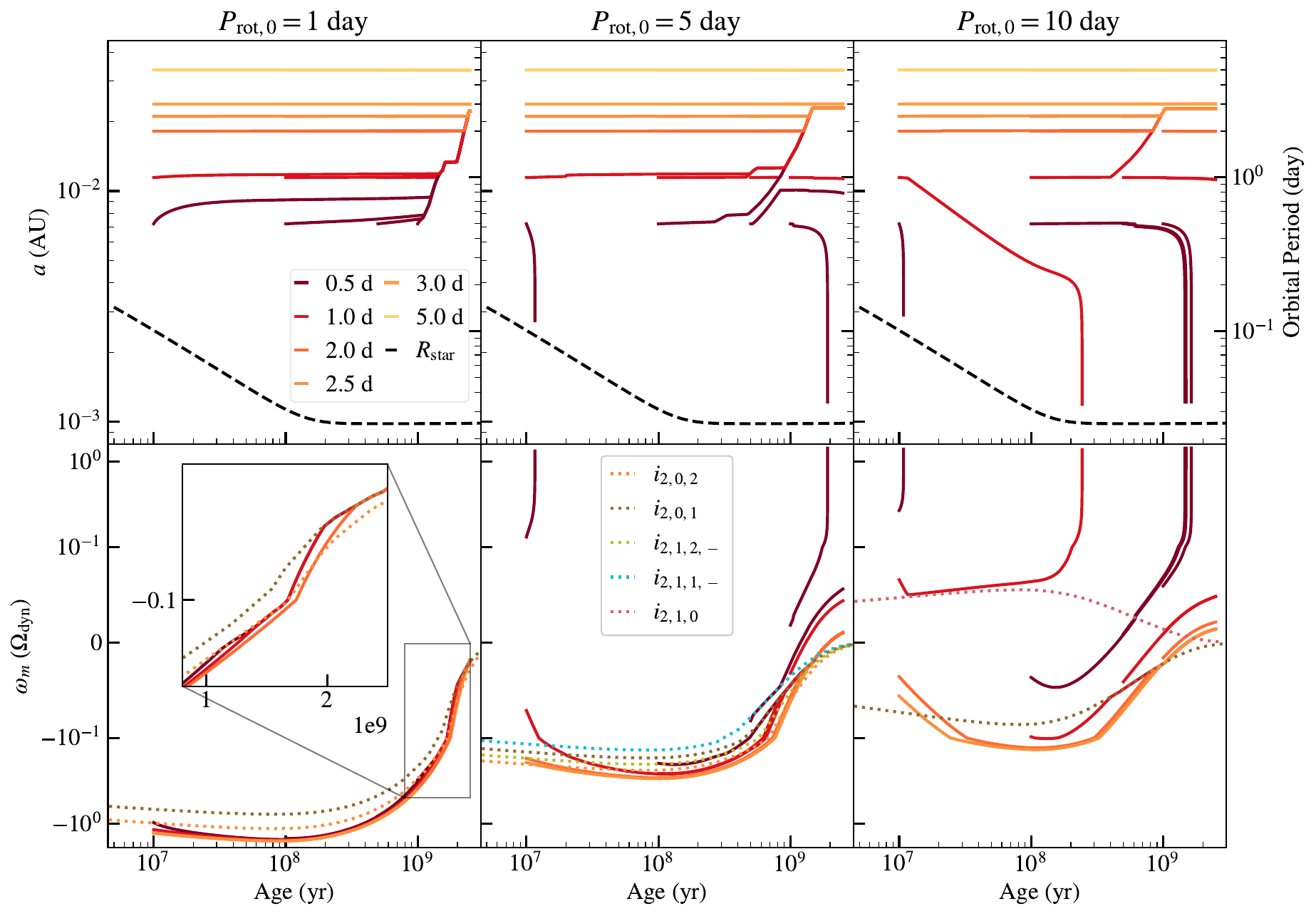}
    \caption{Same as Figure \ref{fig:Mjorbitsummary}, but for a $M_p = M_j$ Jupiter mass companion.
    }
    \label{fig:Mjorbitsummary}
    
\end{figure*}

\section{Orbital evolution due to Stellar Tides}
\label{sec:orbital_results}
%show some summary plots of realistic ones
%maybe show some close orbits for interesting resonance locking behavior

To study the orbital dynamics caused by the frequency-dependent tidal dissipation coupled with stellar evolution as presented in the previous section, we integrate planetary orbits beginning at stellar ages of
% $5$ Myr, $50$ Myr, 
$10$ Myr, $100$ Myr, $0.5$ Gyr, and $1$ Gyr. This spread in initial ages spans scenarios from in-situ or disk-driven formation of close-in planets, in which planets develop $\lesssim 10$ day orbits while within the proto-planetary disk which persists until $\sim 10$ Myr, to dynamically-driven migration, in which planets may reach short-period orbits throughout the stellar lifetime. At each stellar age, we initialize planets at a range of orbital periods from $<1$ day to a few day. We repeat this for an Earth-mass companion ($M_p=M_e$) and a Jupiter-mass companion ($M_p=M_j$), and we also complete the entire procedure for each dissipation grid associated with initial stellar rotation rates $P_{\rm rot,0}=1$, $5$, and $10$ day. For each orbital integration, we assume co-planar, circular orbits and integrate Equations \ref{eq:aevol}--\ref{eq:spinevol} from the respective initial stellar age to a final age of $2.5$ Gyr. The stellar lifetime is of course much longer than this endpoint, but the stellar spin has slowed so drastically by $2.5$ Gyr that our i-mode frequencies approach zero, so beyond this age we expect the f-mode dissipation (i.e., equilibrium tides) to govern the system.

Figure \ref{fig:Meorbitsummary} shows a sample of orbital integrations for $M_p=M_e$ for each initial stellar rotation rate. The top panels show the semi-major axis evolution, while bottom panels show the evolution of the tidal forcing frequency $\omega_m$ and the rotating-frame mode frequencies $\omega_{\alpha}$. For all values of the initial stellar rotation period $P_{\rm rot,0}$, orbits with initial orbital period $P_{\rm orb,0} \gtrsim 1.5$ day experience negligible orbital migration. For these planets that are further away such that the ratio $(R/a)$ is small, $\dot{a}$, which depends on $(R/a)^5$ (Equation \ref{eq:aevol}), remains tiny, and thus the rate of change of the orbital frequency $\dot{\Omega}_o$ does not appreciably affect the evolution of the tidal forcing frequency $\dot{\omega}_m $. Nevertheless, the tidal forcing frequency still changes over time for these orbits due to the stellar spin evolution, but it usually traverses a path where the dissipation away from resonance is too low to affect the orbits. 

For the shortest initial orbital periods shown of $P_{\rm orb,0}=0.5$ and $1$ day, we observe that resonance locking between the tidal forcing frequency and stellar oscillation modes causes significant orbital migration. We first consider the Earth-mass planetary orbits shown in Figure \ref{fig:Meorbitsummary}. For $P_{\rm rot,0} = 1$ day, the star spins faster than all planetary orbits at $P_{\rm orb,0} \gtrsim 0.5$ day so that the planets migrate outward, no matter at what age the integration begins. The planet at $P_{\rm orb,0} =0.5$ day evolves into a brief resonance with the i-mode $i_{2,0,2}$, which is indicated in the bottom panel by the dark blue line overlapping with the dotted pink line (shown more clearly in the inset panel). After it breaks out of resonance with $i_{2,0,2}$ having migrated slightly outward, it enters another resonance with the longest-wavelength retrograde i-mode, $i_{2,0,1}$, and remains in that resonance until $\sim 2$ Gyr. The  planet at $P_{\rm orb,0} = 1$ day experiences the resonance with $i_{2,0,1}$ for a brief interval as well, and its orbit then converges with that of the $P_{\rm orb,0} =0.5$ day planet. By 2.5 Gyr, both of these planets arrive at $P_{\rm orb} >1$ day orbits, regardless of the time of formation.

For dissipation in a more slowly rotating star with $P_{\rm rot,0} = 5$ day, most orbits begin retrograde ($\Omega_{o} < \Omega_s$); however, planets initialized with $P_{\rm orb,0}=0.5$ day at early times $\leq$ 10 Myr are instead prograde ($\Omega_{o} > \Omega_s$). These earlier prograde orbits enter into a resonance lock with the longest-wavelength prograde i-mode $i_{2,1,0}$ that lasts from $\approx 10$--$100$ Myr, causing significant orbital decay to $P_{\rm orb} \sim 0.1$ day. Eventually, between $700$--$800$ Myr, the planet plunges into the star because of the equilibrium tide, which can be understood equivalently as dissipation due to the $\ell=m=2$ f-mode. If these planets arrive at $P_{\rm orb,0}=0.5$ day between 50--400 Myr, they will instead migrate outward in a resonance lock with $i_{2,0,1}$ that breaks at $\sim$1 Gyr; however, if they land at $P_{\rm orb,0}=0.5$ day from $\approx 500$ Myr onward, they are unable to lock into resonance and remain stationary. At $P_{\rm orb,0}=1$ day, orbits remain unaffected by dissipation until $\sim 700$ Myr, at which point they will also lock with  $i_{2,0,1}$ and converge with the $P_{\rm orb,0}=0.5$ day orbits to migrate slightly outward to $P_{\rm orb} >1$ day. However, planets arriving at $P_{\rm orb,0}=1$ day after $\sim$1 Gyr do not undergo migration.

For our slowest set of stellar rotation rates with $P_{\rm rot,0} = 10$ day, planets closer than $P_{\rm orb} \lesssim 1.5$ day will migrate inward until $\sim$100 Myr due to resonance locking with $i_{2,1,0}$. All these orbits converge to $P_{\rm orb} \sim 0.2$ day by $2.5$ Gyr due to equilibrium tidal dissipation. Only planets at $P_{\rm orb,0} = 1$ day initialized between 50--500 Myr migrate outward.
Overall, if Earth-mass planets form short-period orbits of $P_{\rm orb,0} \lesssim 1$ day at around $10$ Myr, then tides tend to pull the planets from this initial position by a few Gyr. However, the final fate (migrating outward, orbital decay, or plunge-in) depends on the initial rotation rate of the star.

Figure \ref{fig:Mjorbitsummary} depicts the orbital evolution for $M_p=M_j$ companions. Jupiter-mass planets suffer stronger tidal dissipation because $\dot{a}$ is proportional to the mass ratio (Equation \ref{eq:aevol}). Thus, even when orbits are not at resonant frequencies, more examples of migration are visible in each panel for $P_{\rm orb,0} < 2$ day. In general, the results for Jupiter mass planets are similar to those for Earth mass planets as they lock into resonance with similar modes at similar times, but with more drastic changes in the semimajor axis. Of note are the Jupiter-mass planets that form at $P_{\rm orb,0} = 0.5$ day, which mostly plunge towards the star in the $P_{\rm rot,0} = 5$ and $10$ day examples. One exception is a planet for the $P_{\rm rot,0} = 5$ day stellar model that migrates outward in a resonance lock with $i_{2,1,1,-}$ at $\approx 500$ Myr. Planets that migrate outward in resonance locks with $i_{2,0,2}$ and $i_{2,0,1}$ tend to approach similar 3 day orbits by 2.5 Gyr, independent of the stellar rotation rate.

\begin{figure}
    \includegraphics[width=\columnwidth]{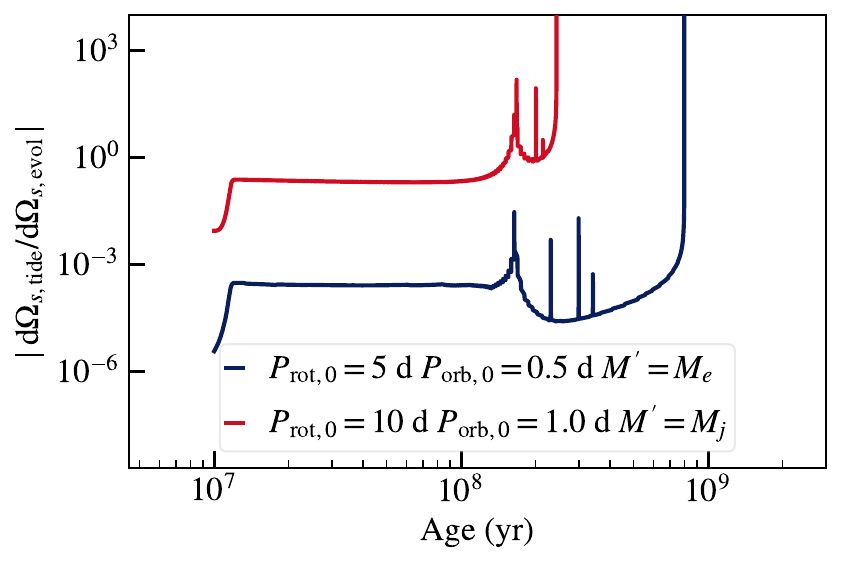}
    \caption{The rate of change of stellar spin from tides due to Equation \ref{eq:spinevol}, $\dot{\Omega}_{s,\rm{tide}}$, compared to that due to stellar evolution (magnetic braking, contraction, or expansion), $\dot{\Omega}_{s,\rm{evol}}$. This is shown for an Earth-mass planet in blue and and a Jupiter-mass planet in red for the models listed in the legend. Note that the nearly vertical red and blue lines at 250 Myr and 800 Myr respectively are due to plunge-in of the planet (see corresponding times for the same models in top right panel of Figure \ref{fig:Mjorbitsummary} and top middle panel of Figure \ref{fig:Meorbitsummary}). 
    }
    \label{fig:spinevol}
    
\end{figure}

As noted in Section \ref{sec:orbits}, we calculate the change in stellar spin without incorporating this feedback into our orbital integrations, since including this effect is computationally prohibitive at this stage. This assumption holds well for Earth-mass planets as long as the planet is not plunging in, as Figure \ref{fig:spinevol} shows. Even when the planet's tidal forcing frequency passes through resonance with mode frequencies, the ratio $|\mathrm{d}\Omega_{s,{\rm tide}}/\mathrm{d}\Omega_{s,{\rm evol}}|$ is small such that the change in stellar spin is dominated by the stellar evolution. In contrast, Jupiter-mass planets exert a stronger tidal torque on the star that significantly affects the stellar spin as the planetary orbit passes through resonances with various modes of the star. This introduces appreciable uncertainty in the ability of certain resonance locks with Jupiter-mass planets to hold for the duration presented in this work, which is discussed further in the next section. Moreover, both Jupiter-mass and Earth-mass planets during plunge-in traverse nearly vertical lines in Figure \ref{fig:spinevol} and attain large values of $|\mathrm{d}\Omega_{s,{\rm tide}}/\mathrm{d}\Omega_{s,{\rm evol}}|$. This trajectory indicates that both types of planets can greatly increase the stellar spin as their orbits decay into the star, potentially leading to tidal synchronization of the spin and orbit.

\subsection{Resonance locking}
\label{sec:reslocks}
In our orbital evolution calculations, we observe resonance locks that tend to occur between a few--100 Myr for inwardly-migrating orbits, as well as later at a few hundred Myr--1 Gyr outwardly-migrating orbits. To understand why resonance locks occur, we first note the conditions for resonance locking. The tidal forcing frequency must be nearly equal to the frequency of a mode: $\omega_m \simeq \omega_{\alpha}$. For the lock to be maintained, the time derivatives must remain equal as well: $\dot{\omega}_m \simeq \dot{\omega}_{\alpha}$. Once the resonance lock begins, the companion orbit will remain at a stable fixed point where this condition is true, unless something occurs to break the lock. The orbital migration timescale during the resonance lock follows the evolutionary timescale of the stellar oscillation mode: as the mode frequency changes over time, the tidal forcing frequency follows closely \citep{fuller2016}. 

Recalling that $\omega_m=m(\Omega_o-\Omega_s) \Rightarrow \dot{\omega}_m = m(\dot{\Omega}_o-\dot{\Omega}_s)$, we see that the ability to resonance lock depends on the interplay between the stellar spin evolution and the orbital evolution. In this work, we do not incorporate the feedback of the tidal dissipation onto the stellar spin, so $\dot{\Omega}_s$ is given by the slope of the curves in Figure \ref{fig:starprops}. Tidal dissipation drives the change in semimajor axis $\dot{a}$, and consequently $\dot{\Omega}_o$, through Equation \ref{eq:aevol}. Given that the planet is close enough so that $(R/a)^5$ is non-negligible, then the term $\dot{\Omega}_o$ will begin to contribute when the dissipation from $\mathrm{Im}\left[k_{2,2}^2\right]$ becomes large enough. This will occur near normal mode frequencies where the dissipation rises sharply into peaks.

We can understand migration driven by resonance locking from the condition $\dot{\omega}_{\alpha} \simeq \dot{\omega}_m$. Inertial modes have $\omega_\alpha \simeq c \Omega_s$, where $-2\leq c \leq 2$ is nearly constant over time. Hence a resonance lock with an inertial mode requires
\begin{equation}
\label{eq:reslocki}
    \dot{\Omega}_o \simeq \frac{2+c}{2} \, \dot{\Omega}_s \, .
\end{equation}
In a resonance lock with an inertial mode, the planet can only migrate inwards if the star is spinning up, and it can only migrate outwards if the star is spinning down, such that $\dot{\Omega}_o$ and $\dot{\Omega}_s$ have the same sign.

\begin{figure}

    \includegraphics[width=\columnwidth]{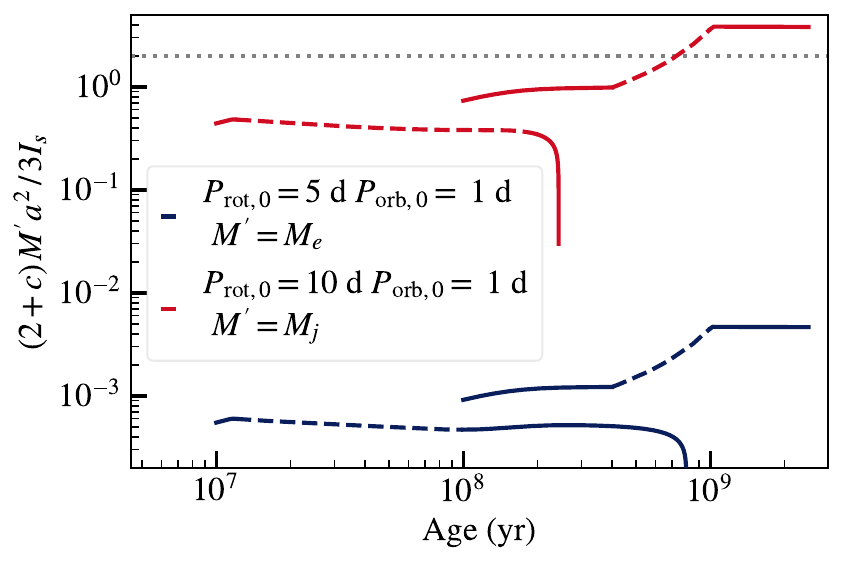}
    \caption{The value of the term $(2+c) M' a^2/3 I_s$ in Equation \ref{eq:reslockistar} during resonance locking shown for the models listed in the legend. Orbits evolving from $10$ Myr that migrate inward and from $100$ Myr that migrate outward, as well as both Earth-mass (blue) and Jupiter-mass (red) planets, are shown. The resonance lock duration is shown as dashed sections in each curve. The dotted gray line denotes where $(2+c) M' a^2/3 I_s = 2$. Above this value, resonance locks will break due to tidal torque on the stellar spin, which was not included in these models.
    }
    \label{fig:reslockspinevol}
    
\end{figure}

\begin{figure*}
    % \centering
    \includegraphics[width=0.5\textwidth]{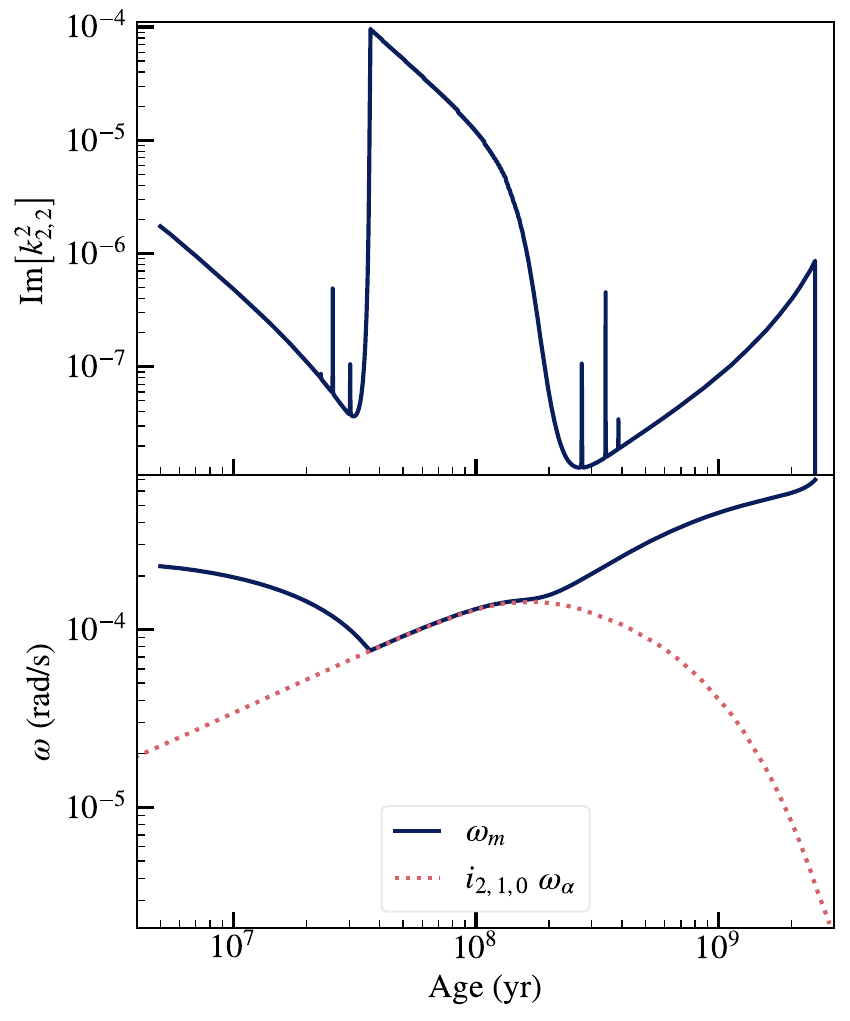}
    \includegraphics[width=0.5\textwidth]{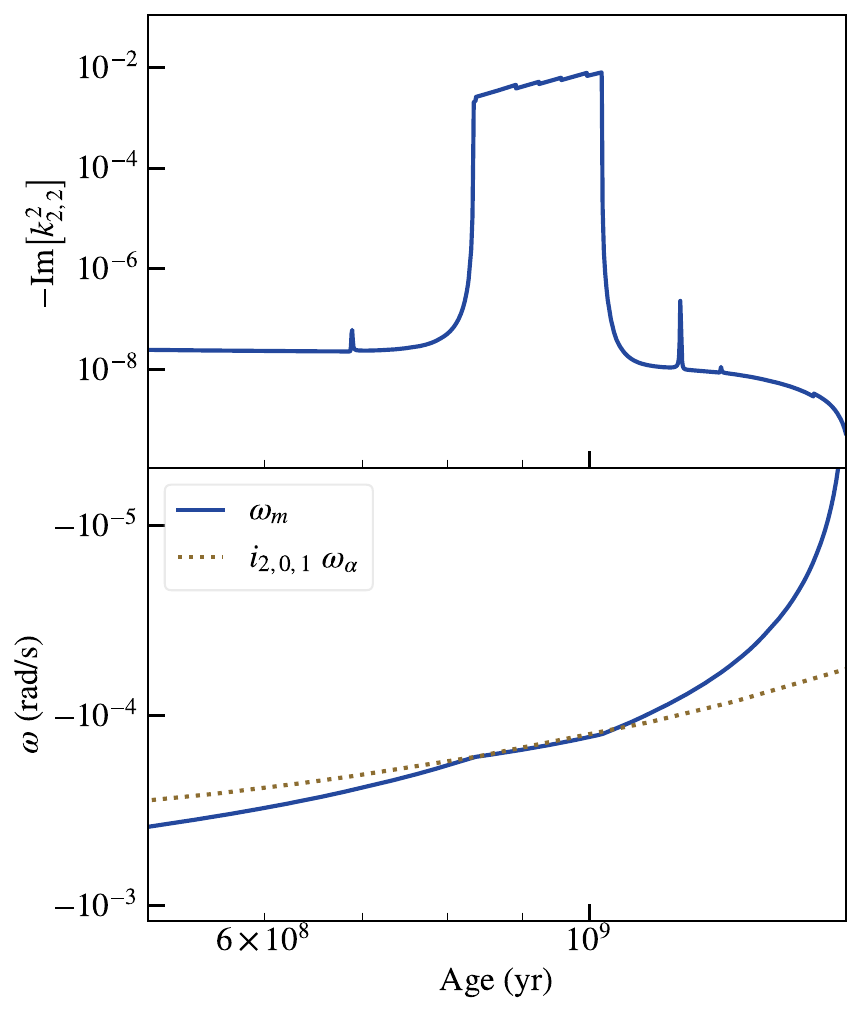}
    \caption{Examples of resonance locking for Earth-mass exoplanets %with dissipation (using the correction to viscosity under rotation, Equation \ref{eq:Rvisc}))
    from the $P_{\rm rot,0}=10$ day initial condition, for initial orbital periods $0.5$ (left) and $1$ day (right). The top panels show the time evolution of the imaginary part of the Love number $k_{2,2}^2$ (positive for prograde orbits $\omega_m >0$, negative for retrograde orbits $\omega_m <0$). The bottom panels show the tidal forcing frequency $\omega_m$ and the rotating-frame mode frequency $\omega_{\alpha}$ for the i-mode in the legend.
    %and the bottom panel shows the rate of change of $\omega_m$ as well as the time derivative of each of the two terms contributing to $\omega_m$.
    The left column shows an inwardly migrating resonance lock with a prograde mode $\omega_{\alpha} > 0$, and the right exemplifies an outwardly migrating resonance lock with a retrograde mode $\omega_{\alpha} < 0$.
    }
    % k22: from 9e-5 to 4e-5 on left; from 2e-3 to 6e-3 on right
    % da/dt: from 4e-5 to 1e-6 on left; 6e-6 on right
    % r/a: from 4.5 on left to 11 on right (this is to power of 5). ratio of (r/a)^5 for each is about 90, accounts for differeence in k22 of about 100 for similar da/dt
    \label{fig:reslocksummary}
    %both are Mearth
\end{figure*}

\begin{figure}
    % \centering

    % \includegraphics[width=\columnwidth]{Prot10_Porb0.5_reslock_k22vsomm.pdf}
    \includegraphics[width=\columnwidth]{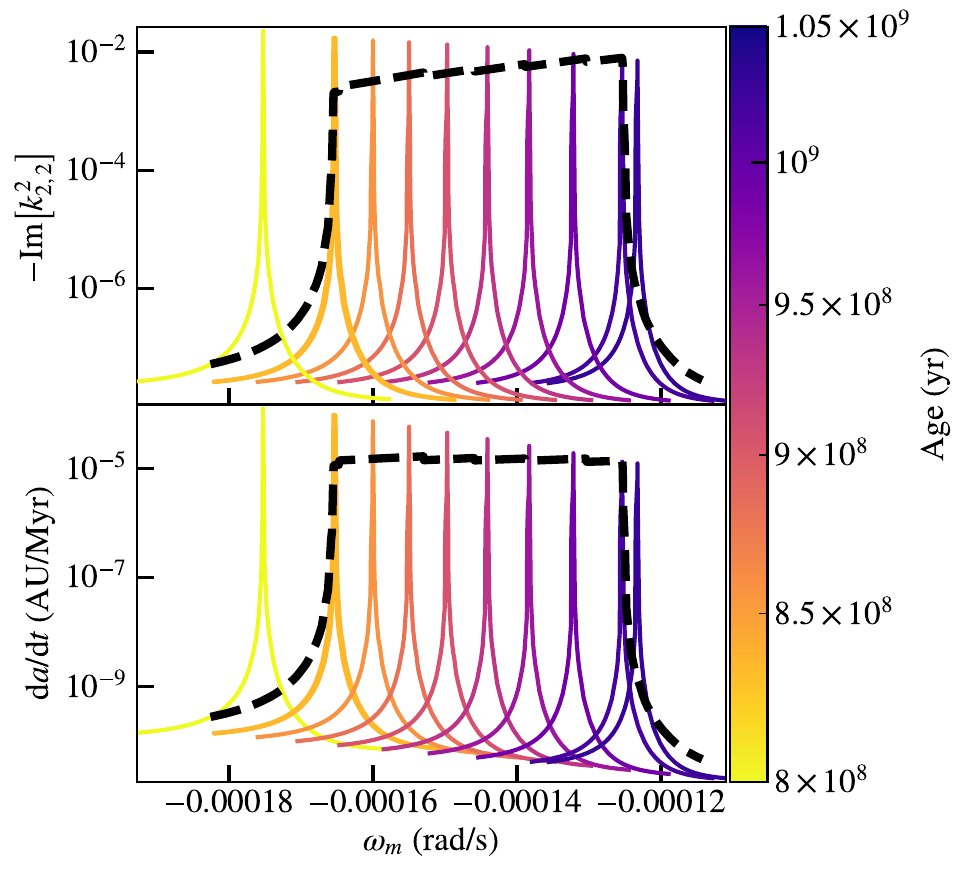}
    
    \caption{The imaginary part of the Love number (top panel) and the rate of change of the semimajor axis (bottom panel) versus tidal forcing frequency $\omega_m$. For this example of a retrograde resonance lock, $\mathrm{Im}[k_{2,2}^2] <0$ so $-\mathrm{Im}[k_{2,2}^2]$ is plotted.
    %On the left, the semimajor axis derivative is negative, and on the right, it is positive.
    Black dashed lines show the evolution of each quantity for the planetary orbit.
    %for the evolution of resonance locks as shown in Figure \ref{fig:reslockk22} (left is $P_{\rm rot,0} = 10$ day, $P_{\rm orb,0} = 0.5$ day and right is $P_{\rm rot,0} = 10$ day, $P_{\rm orb,0} = 1$ day, both for $M_p=M_e$),
    The peaked lines show the shape of the dissipation curve for a subset of times in the orbital evolution, shaded by the stellar age as shown in the colorbar. The dissipation is sharply peaked due to the resonant mode $i_{2,0,1}$, and the peak center moves to the right as the mode frequency $\omega_{\alpha}$ evolves in time. The peak height decreases with time as the star spins down and the mode dissipation weakens. The resonance lock breaks when the dashed black line exceeds the height of the resonance peak.
    }
    \label{fig:reslockk22}
    
\end{figure}

We can further decompose the spin evolution into a component from stellar evolution, $\dot{\Omega}_{s,{\rm evol}}$, and a component due to tidal torques, $\dot{\Omega}_{s,{\rm tide}}$. The latter can be related to the orbital evolution and  $\dot{\Omega}_{s,{\rm evol}}$ via equations \ref{eq:aevol} and \ref{eq:spinevol}. The orbital evolution during the resonance lock can then be written as
\begin{equation}
\label{eq:reslockistar}
    \dot{\Omega}_o \simeq (2+c) \bigg[2 - \frac{(2+c) M' a^2}{3 I_s} \bigg]^{-1} \, \dot{\Omega}_{s,{\rm evol}} \, .
\end{equation}
The second term in brackets accounts for the back-reaction of the tidal torque on the star's spin, which we presently ignore in our orbital evolution calculations. It becomes important when the planet's orbital angular momentum is comparable to that of the star. In particular, as long as $(2+c) M' a^2/3 I_s <2$, then $\dot{\Omega}_o$ and $\dot{\Omega}_{s,{\rm evol}}$ have the same sign and resonance locking therefore continues to be possible.

In Figure \ref{fig:reslockspinevol}, the second term in brackets in Equation \ref{eq:reslockistar} is shown for some orbits that experience resonance locking. Equation \ref{eq:reslockistar} is only valid throughout the duration of the resonance lock, which lasts for $\sim $100 Myr in each orbit shown (dashed lines). At the start of each resonance lock, $\dot{\Omega}_o $ and $\dot{\Omega}_s $ are required to have the same sign in order to enter the resonance lock and satisfy Equations \ref{eq:reslocki} and \ref{eq:reslockistar}. Comparing the right hand sides of Equations \ref{eq:reslocki} and \ref{eq:reslockistar} shows that this is equivalent to $\dot{\Omega}_o $ and $\dot{\Omega}_{s,{\rm evol}} $ having the same sign in our models, as $(2+c) M' a^2/3 I_s <2$ at the start of each resonance lock. For the majority of the models shown, the term $(2+c) M' a^2/3 I_s$ never exceeds 2, so the resonance lock may continue.

For the outwardly migrating Jupiter-mass planet whose orbital evolution begins at 50 Myr, the term $(2+c) M' a^2/3 I_s$ crosses $2$ (dotted gray line) during the resonance lock at 700 Myr, when the planet migrates past $P_{\rm orb} \approx 2$ day.
Once $(2+c) M' a^2/3 I_s > 2$, the tidal torque on the star will become important and cause the right-hand side of Equation \ref{eq:reslockistar} to switch sign. This implies that the star would have to spin up due to stellar evolution, $\dot{\Omega}_{s,\rm{evol}} >0$, in order to maintain the resonance lock condition. Since the star is still spinning down at this point in our stellar models due to magnetic braking, we know the resonance lock condition should actually be broken at this point.
% Comparing Equations \ref{eq:reslocki} and \ref{eq:reslockistar} shows that $\dot{\Omega}_s$ will switch sign as well. At this point, we know the resonance lock condition should actually be broken because $\dot{\Omega}_o $ and $\dot{\Omega}_s $ now have opposite sign. 
As a result, we predict that including the back-reaction of the tidal torque on the stellar rotation will cause resonance locks with outwardly-migrating Jupiter-mass planets to break once they reach $P_{\rm orb} \approx 2$ day orbits. In fact, the tidal torque could synchronize the stellar spin with the orbits of these outwardly-migrating Jupiter-mass planets.

For Earth mass planets whose resonance locking is treated accurately in our models, Figure \ref{fig:reslocksummary} shows some examples of the dissipation and frequency evolution during the lock. The left panel exemplifies a resonance lock with a prograde i-mode. The bottom panel shows that the tidal forcing frequency approaches the value of $\omega_{\alpha}$, which is also changing with time. The relevant mode frequency increases as the star spins up until $\sim 100$ Myr, then decreases with the stellar spin-down thereafter. 
Once $\omega_m \simeq \omega_{\alpha}$, the value of $\mathrm{Im}\left[k_{2,2}^2\right]$ jumps by 3 orders of magnitude as the planetary orbit encounters the peak in dissipation due to resonance with $i_{2,1,0}$.
On the right, Figure \ref{fig:reslocksummary} depicts a resonance lock for retrograde orbits with $\omega_m < 0$ that occurs around a few 100 Myr--1 Gyr for a subset of initial conditions. 
These planets lock into resonance with $i_{2,0,1}$.

%explain why it breaks
In both cases, the resonance locks break after $\sim$ $100$s of Myr, which occurs when the stable fixed point of the resonance lock can no longer be maintained. For the prograde orbits on the left, the resonance lock is initially maintained as the mode frequency and tidal forcing frequency are both increasing. However, once the stellar rotation switches from spinning up to spinning down, the i-mode frequency begins to decrease and the resonant mode peak moves to decreasing forcing frequencies. The orbit thus loses the stable fixed point.

In the case of the retrograde resonance lock, the star is continuously spinning down by the time the lock begins at a few $100$ Myr, so the lock breaks for a different reason.
Figure \ref{fig:reslockk22} depicts the evolution of the relevant resonant mode peak for retrograde orbits shown in Figure \ref{fig:reslocksummary}, in terms of the value of $\mathrm{Im}\left[k_{2,2}^2\right]$ (top) and the rate of change of the semimajor axis (bottom). The black dashed line shows the value of each quantity achieved by the planetary orbit, which in the regime of heightened dissipation can be thought of as the necessary value to maintain the fixed point.
Since the peaks in both panels become shorter as the stellar rotation rate decreases, the maximum possible dissipation diminishes significantly over time. Furthermore, the value of $\mathrm{Im}\left[k_{2,2}^2\right]$ required to maintain the fixed point actually increases with time, since more dissipation is required to sustain the necessary $\dot{a}$ as the planet migrates away. Thus, the maximum dissipation available from the resonant peak is eventually unable to sustain the torque required for the fixed point.  

For the $P_{\rm orb,0} = 0.5$--2 day orbits of Earth-mass planets shown in Figure \ref{fig:Meorbitsummary}, the only resonance locks occur for $i_{2,0,1}$, $i_{2,1,0}$, and $i_{2,0,2}$. On the other hand, for much closer orbits $\lesssim 0.5$ day, we find that nearly every stellar oscillation mode is able to sustain resonance locks, given amenable initial conditions such that the tidal forcing frequency of the orbit can intersect with the mode frequency. 
% Figure \ref{fig:Meorbitsummary2} demonstrates the range in behavior for an Earth-mass planet: for $P_{\rm rot,0} = 1$ day, planets either plunge rapidly toward the star or migrate away by $2.5$ Gyr due to resonance locks with a variety of i-modes; whereas for $P_{\rm rot,0} = 5$ or $10$ day, most planets either plunge quickly or migrate inward due to resonance locks with a few prograde i-modes before plunging. Planets that have not yet experienced much orbital decay by $2.5$ Gyr are expected to continue to migrate inward due to off-resonant dissipation by the prograde $\ell=m=2$ f-mode and eventually fall towards the star. In any case, we expect a dearth of planets in this short-period regime due to a variety of resonance locks in combination with strong non-resonant dissipation near $f_{2,2,+}$.

\subsubsection{Viscosity and non-linearity}
\label{sec:orbvisc}
As discussed in Section \ref{sec:viscdiss}, the off-resonant dissipation is orders of magnitude larger when calculating tidal dissipation using the viscosity independent of Rossby number $\nu_{\mathrm{NR}}$ (Equation \ref{eq:NRvisc}). Furthermore, the resonant peaks associated with stellar oscillation modes are both wider and shorter. For the resulting orbital evolution, we find that the effect of using the larger viscosity $\nu_{\rm NR}$ is an increased prevalence of planets plunging towards the star due to non-resonant dissipation. In addition, there are fewer initial conditions $P_{\rm orb,0}$ for which resonance locking occurs. 
Compared to the fiducial models shown in Figure \ref{fig:Meorbitsummary}, we find that the prograde orbits decay on a shorter timescale, and the $P_{\rm orb,0} \gtrsim 1$ day orbits do not catch into resonance. Since the resonant mode peaks are shorter for this higher viscosity, $P_{\rm orb,0}=$1 day orbits are now too far away for the resonant peak in the dissipation to be able to sustain the requisite torque for a resonance lock.

If the convective viscosity is much larger \citep[or if the interaction of convective and tidal flows produces enhanced dissipation that is not well-described with a convective viscosity; ][]{terquem2021}, then equilibrium tidal dissipation would likely dominate the evolution rather than locks with inertial modes. However, if convective viscosity is smaller via the \cite{goldreich1977} prescription as demonstrated by recent work \citep{duguid2020a,duguid2020b}, then equilibrium tidal dissipation will be less important. Tidal migration in that case is likely to be dominated by resonance locks, even for fairly massive planets. We leave the consideration of different prescriptions for the viscosity in the regime of rapid tides to future work.

Our study assumes that modes are linear, which we find holds true even during the resonance locks shown in Figures \ref{fig:Meorbitsummary}--\ref{fig:Mjorbitsummary}. Though resonance locking drives modes to larger amplitudes, typically the amplitude (Equation \ref{eq:amps}) remains small. In the case of Earth-mass planets, the mass ratio is small enough to ensure small mode amplitudes throughout resonance locks. Even for Jupiter-mass planets, in many cases the tidal overlap integral $Q_{\ell,m}^{\alpha}$ is small because the resonance lock occurs while the star is rotating slowly; during other resonance locks, either the frequency separation $\omega_{\alpha}-\omega_m$ or the mode damping rate $\gamma_{\alpha}$ is relatively large. Overall, we estimate mode amplitudes $\xi$ on the order of $\sim 10^{-5}\text{--}10^{-4}$, and wavenumbers $k\sim \sqrt{n_1^2+n_2^2+m^2} \lesssim 4$, so that the nonlinearity measure $k\xi < 1$. Whether weakly non-linear dissipation (e.g., \citealt{weinberg2012}) can increase tidal dissipation or prevent resonance locking is unclear for fully convective stars, and should be studied in future work.

\section{Discussion}
\label{sec:discussion}

\subsection{Comparison with Observed Systems}

The occurrence rate of Earth-like planets around M-dwarfs in the Kepler sample is less than a few percent for short orbital periods $P_{\rm orb,0} \lesssim 1$ day \citep{Mulders_2015,hsu2020}, where our models predict migration due to resonance locking. The story appears to be similar for TESS, as recent analyses find only a few planets with Earth-like radii at $\lesssim 1$ day orbits out of the dozens of planets orbiting M-dwarfs in their sample \citep{rodriguez2023}. The vast majority of planets lie at a few--$10$ day orbits, where negligible migration occurs in our fully convective stellar models. Our results indicate that a dearth of Earth-mass planets at $P_{\rm orb} \lesssim 1$ day around fully-convective M-dwarfs is expected due to migration induced by resonance locks. For Earth-mass planets at $P_{\rm orb,0} \lesssim 1$ day, planets orbiting slower than the stellar spin tend to migrate out to similar $\sim$1.5 day orbits, whereas planetary orbits rotating faster than the stellar spin tend to decay towards the star. For Jupiter-mass planets, we expect few planets out to $\sim$3 days as a result of migration due to resonance locking.  
%compare the initial conditions to occurrence rate of planets e.g. mulders for Kepler survey. e.g. the 0.5 day stars, while rare, are more common for M dwarfs at least. but less than a few percent will migrate then according to the mulders,hsu paper for kepler.

By initializing our orbital evolutionary calculations at different times, we are able to explore how our predicted outcomes vary for the timescales associated with different theories of the origin of short-period planets. For in-situ planet formation or disk migration theories, changes to the orbital separation are dominated by the proto-planetary disk until the disk dissipates at a few--$10$ Myr \citep{dawson2018}. High-eccentricity tidal migration may operate on a large range of timescales, as it is highly sensitive to the initial conditions of the underlying planet-planet interactions. For instance, the secular interactions and the ensuing tidal migration to short orbital periods may span anywhere from $\sim$Myr to tens of Gyr \citep{dawson2018}.  

For the fastest-spinning $P_{\rm rot,0}=1$ day model, the Earth-mass planets will experience negligible migration until the resonance lock at $\sim$1-2 Gyr causes planets at $P_{\rm orb,0} < 1$ day to migrate outward. This is true during the first Gyr of the stellar evolution, no matter what time the orbital integration begins. As a result, we predict the same behavior regardless of when the planet reached its short orbital period, as long as it occurs before $\sim 1$ Gyr.

In our slower-rotating stellar models, an assortment of behaviors may ensue depending on the formation timescale of short-period planets. For instance, in the $P_{\rm rot,0}=5$ day set of integrations, formation theories placing planets at $P_{\rm orb,0} \lesssim 1.5$ day at stellar ages $\lesssim 50$ Myr are susceptible to resonance locking, either immediately beginning inward migration or experiencing delayed outward migration at a few $100$ Myr--$1$ Gyr. However, if the Earth-mass planet is driven to short orbital periods at $\gtrsim 50$ Myr, the orbit may not be affected. For example, evolving a $P_{\rm orb,0} = 0.5$ day planet from $500$ Myr in the $P_{\rm rot,0}=5$ day model is too late for a resonance lock. In the $P_{\rm rot,0}=10$ day model, the $P_{\rm orb,0} = 0.5$ day evolution initialized at $100$ Myr is unable to lock into resonance, though the same evolution initialized at $10$ Myr migrates inward due to resonance locking. Similar variations in behavior for different initial conditions can be observed for the $P_{\rm orb,0}=1$--$1.5$ day curves in Figures \ref{fig:Meorbitsummary}.

In general, the dependence of evolutionary outcomes on the different timescales of short-period planet formation can be attributed to the variation in initial tidal forcing frequency $\omega_m$ as the stellar rotation rate changes over time. This aspect of the stellar evolution may prevent the planet from achieving the same resonance lock and enhanced migration that occurs for another initial condition.  Figure \ref{fig:Mjorbitsummary} presents an analogous range in evolutionary outcomes that depend on the potential formation timescales for short-period Jupiter-mass planets, if they are found around low-mass M-dwarfs. Nevertheless, the caveat discussed in Section \ref{sec:reslocks} for outwardly-migrating Jupiter-mass planets remains relevant here.

\begin{figure*}
    % \centering

    \includegraphics[width=0.5\textwidth]{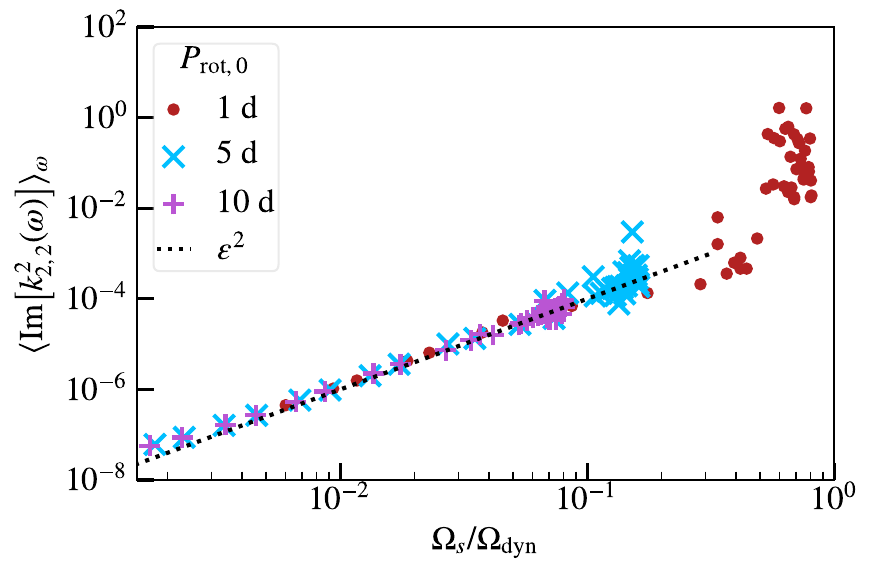}
    \includegraphics[width=0.5\textwidth]{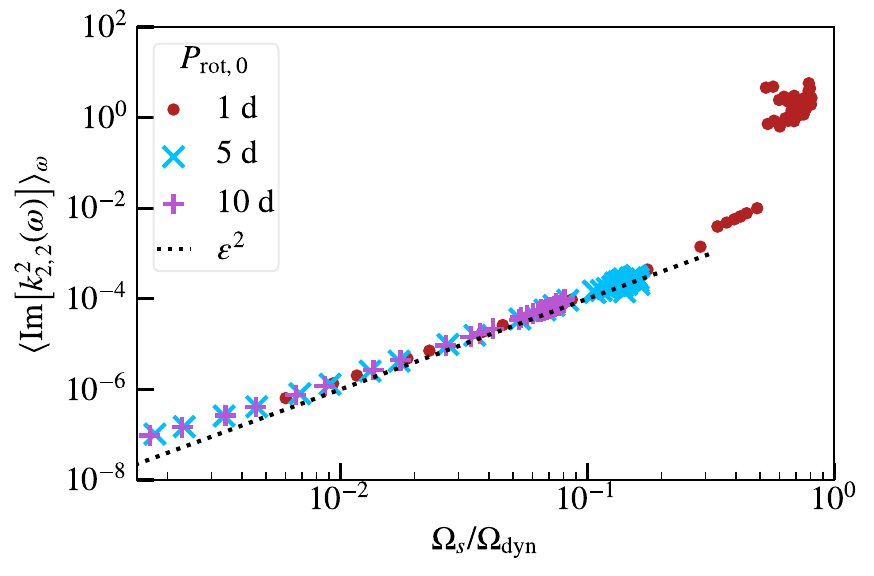}   
    \caption{Frequency-averaged dissipation $\langle \mathrm{Im}\left[k_{2,2}^2\right]\rangle_{\omega} = \int_{-2\Omega_s}^{2\Omega_s} \mathrm{Im}\left[k_{2,2}^2\right] d\omega/\omega$ as in \cite{ogilvie2013}. The left panel shows the result for dissipation using rotationally modified viscosity $\nu_{\mathrm{R}}$ (Equation \ref{eq:Rvisc}), and the right shows the result with the unmodified viscosity $\nu_{\mathrm{NR}}$ (Equation \ref{eq:NRvisc}). For comparison, the scaling in black matches the prescription in \cite{ogilvie2013} of $\langle \mathrm{Im}\left[k_{2,2}^2\right]\rangle_{\omega}\propto \epsilon^2=(\Omega_s/\Omega_{\rm dyn})^2$.
    }
    \label{fig:freqavg}
    
\end{figure*}

The majority of the migration illustrated in this work occurs before a few Gyr, so observing evidence of migration in the orbital architectures of exoplanets would be facilitated by the detection of young M-dwarf systems. 
In our predictions, planets with $P_{\rm orb} \lesssim 1$ day that move outward to larger $P_{\rm orb}$ often begin experiencing migration at $\sim \! 1$ Gyr. For older M-dwarf systems found at a few Gyr, these planets will have already reached larger separations and vacated the region of short-period orbits. If detected in future observations, 
e.g. from TESS or the Roman Space Telescope, short-period planets around M-dwarfs younger than $\lesssim \! 1$ Gyr likely have not yet migrated outward. As a result, we predict short-period planets to be more common around young M-dwarf systems than older M-dwarf systems. 

Looking at the NASA Exoplanet Archive, we note that there are not many young, low-mass stars hosting short-period planets due to the  challenges in observing planets around active young stars. Nevertheless, some young, rapidly rotating stars hosting planets at short periods such as TOI 540b \citep{ment2021} and K2-25b \citep{thao2020,stefansson2020} are interesting candidates to undergo tidal evolution in the context of our models. In the future, we will be able to better test our model's predictions given more observations of planets around young stars.
Nevertheless, to make a definitive prediction for evidence of tidal migration in the period distribution of observed exoplanet systems would require more extensive modeling than done in this work. The period distribution is shaped by combination of the physics of planet formation and the effect of tides, so it will be necessary to model both of these aspects in tandem (e.g., \citealt{lee2017}) order to make more robust predictions in the future. 

In exoplanet systems with more massive host stars, the star is not fully convective. Thus, the dynamical tide likely comprises gravito-inertial waves instead of the pure inertial modes examined in this work. For young, rapidly-rotating solar type stars with planetary companions at $\lesssim 5$ day, for instance the host stars TOI 942 \citep{wirth2021} and WASP 25 \citep{bonomo2017}, we do expect similar orbital dynamics due to gravito-inertial modes relative to what we discuss in this work. However, the denser spectrum of gravito-inertial modes at small frequencies will complicate the picture by involving contributions to the tidal dissipation from a larger number of normal modes. We anticipate that the rich spectrum of mode peaks will warrant detailed exploration of resonance locking effects in these systems.

\subsection{Comparison to frequency-averaged dissipation}
Models of similar star-planet systems in the past have relied upon a prescription for the frequency-averaged dissipation due to the dynamical tide. For instance, \cite{gallet2017} modeled dynamical tides in stars between $0.3$--$1.4\, M_{\odot}$ based on a prescription for frequency-averaged dissipation from \cite{ogilvie2013}:
\begin{align}
    \label{eq:imk2ogilvie}
    \langle \mathrm{Im}\left[k_{2,2}^2(\omega)\right] \rangle_{\omega}& = \int_{-2\Omega_s}^{2\Omega_s} \mathrm{Im}\left[k_{2,2}^2(\omega)\right]  \frac{d\omega}{\omega}\\
    \notag
    &=\frac{100\pi}{63} \epsilon^2 \frac{\alpha^5}{1-\alpha^5},
\end{align}
where $\epsilon = \Omega_s/\Omega_{\rm dyn}$ and $\alpha =R_c/R$ is the fractional size of the radiative core. In their models, no tidal dissipation occurred during periods of the evolution where the star was fully convective because of their prescription's dependence on $\alpha$. As a result, for our models where $\alpha=0$, equation \ref{eq:imk2ogilvie} underestimates the amount of tidal dissipation.

We can also compare the how dissipation scales with  $\epsilon$.
Figure \ref{fig:freqavg} shows the value of $\langle \mathrm{Im}\left[k_{2,2}^2(\omega)\right] \rangle_{\omega}$ in our models, with the fiducial models using $\nu_\mathrm{R}$ on the left, and the models with $\nu_{\mathrm{NR}}$ on the right. We see the same scaling of $\epsilon^2$ in our models up to values of $\epsilon\gtrsim 0.2$, where scatter begins to accumulate around the relation. This is not surprising, as the scaling was derived for slowly rotating bodies where the f-modes are well separated from the i-modes in frequency, assuming a constant dynamic viscosity \citep{ogilvie2013}. The $P_{\rm rot,0} = 1$ day models form a group of outliers as they encounter avoided crossings between f-modes and i-modes for $\epsilon \gtrsim 0.6$, such that the frequency range of $[-2\Omega_s,2\Omega_s]$ includes the large f-mode dissipation. This strongly amplifies the value of the frequency-averaged dissipation. We also note that the models with $\nu_{\mathrm{NR}}$ follow the trend more closely than models using $\nu_\mathrm{R}$ because the variation of viscosity with Rossby number within the star introduces more scatter.

Compared to Figure 4 of \cite{gallet2017}, our frequency-averaged dissipation values, which represent $\alpha=0$ in the context of their work, span a similar order-of-magnitude range of $\langle \mathrm{Im}\left[k_{2,2}^2(\omega)\right] \rangle_{\omega} \approx 10^{-8}$--$10^{-4}$  for $\epsilon \lesssim 0.1$. In their higher-mass models, stars achieve this level of dissipation with core sizes $\alpha \approx 0.2$--$0.8$ throughout the evolution. Their most similar model in terms of stellar mass is an $0.3\, M_{\odot}$ model which only hosts a radiative core for a few tens of Myr and correspondingly only exhibits dissipation during that period. In contrast, for our low-mass M-dwarfs that are fully convective, we predict frequency-averaged dissipation at the same level as their solar-mass models throughout the stellar lifetime. Furthermore, the goal of this work was to instead use a frequency-dependent tidal dissipation in our calculations, which reveal a variety of scenarios involving resonance locking that are not captured by an average over frequency.

\begin{figure}
    % \centering
    \includegraphics[width=\columnwidth]{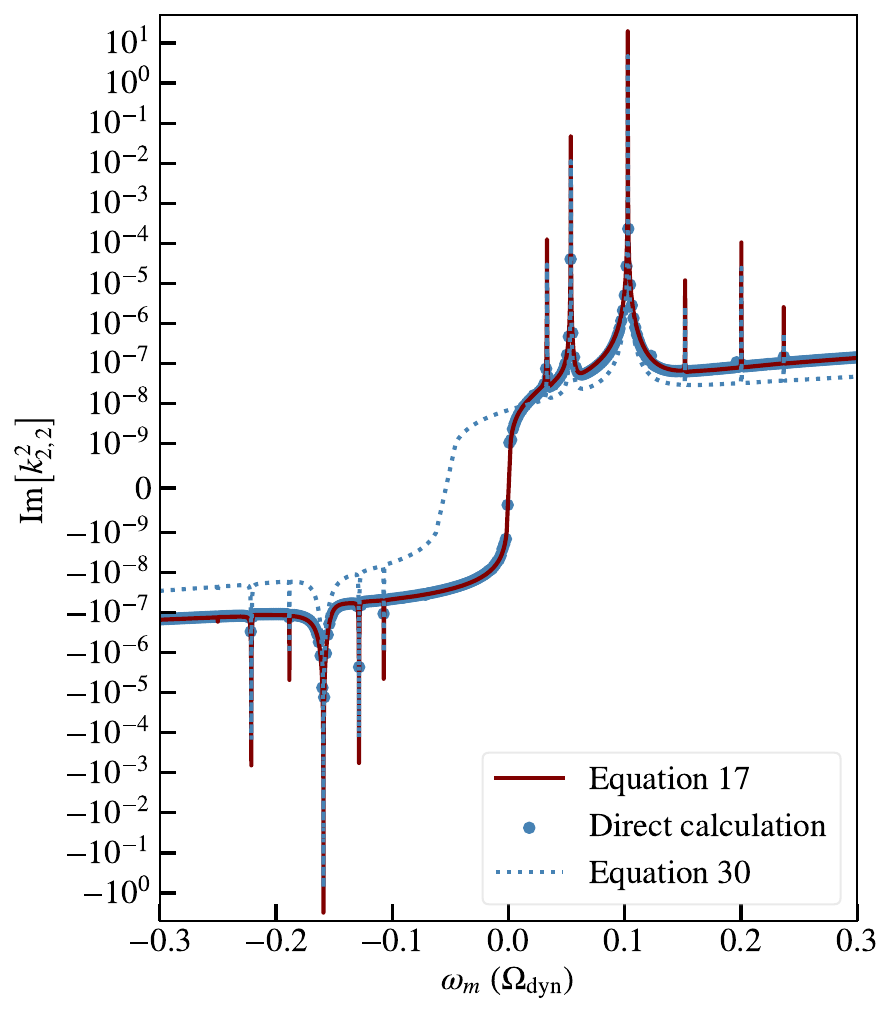}
    \caption{Comparison of different methods for computing the Love number, shown for a $P_{\rm rot,0}=5$ day, $\Omega_s/\Omega_{\rm dyn} = 0.150$ model. The red solid curve shows the result of using Equations \ref{eq:D}--\ref{eq:Dklm} to infer $\mathrm{Im}\left[k_{2,2}^2\right]$, which includes dissipative mode coupling terms. The blue dots show the results from a direct forced calculation, which are nearly indistinguishable from the red line. In contrast, the dotted blue line shows the result from a sum over normal modes without dissipative coupling terms (Equation \ref{eq:klmsum}), which deviates significantly from the other two methods away from resonance.
    }
    \label{fig:comparemethods}
    
\end{figure}

\subsection{Total dissipation vs. individual mode damping}
\label{sec:modamp}
Multiple approaches to calculating dissipation from the dynamical tide exist in the literature, primarily by direct calculation or by modal decomposition. The direct calculation refers to numerically solving the linearized fluid dynamics equations to find the perturbed response of a fluid under the influence of a tidal potential with forcing frequency $\omega_m$ \citep[see, e.g.,][]{Dewberry2023,ogilvie2013}{}{}. To recover the spectrum of $\mathrm{Im}\left[k_{2,2}^2\right]$ across forcing frequency, the calculation must be repeated at each value of $\omega_m$. Other works use an expansion over oscillation modes, which constitutes the theoretical framework of this work \citep[e.g.,][]{press1977,Kumar1995,lai1997,fuller2012,fuller2017}. This popular approach allows for a less numerically expensive analysis that readily captures resonances with normal modes of the star.

However, \cite{townsend2023} find a discrepancy between the two methods in the context of modeling secular tidal torques in binary star systems. They attribute the issue to a subtlety about expanding over normal modes: it is insufficient to consider each mode as contributing independent coefficients characterized by individual mode damping rates $\gamma_{\alpha}$. For instance, works using modal decomposition typically find the Love number from the following sum \citep[e.g.,][]{Dewberry2023}{}{}:
\begin{equation}
    \label{eq:klmsum}
    k_{\ell,m}^n = \frac{2\pi}{2\ell+1}\sum_{\alpha}\frac{Q_{\ell m}^\alpha Q_{n m}^\alpha}{\epsilon_{\alpha}(\omega_{\alpha}-\omega_m-i\gamma_{\alpha})}.
\end{equation}
This method employs Equation \ref{eq:veldamped} to compute the mode amplitudes.
% , and dissipation is computed from a sum over modes.

Though we also use Equation \ref{eq:veldamped} to calculate the mode amplitudes, we obtain coefficients in the damping integral (Equation \ref{eq:D}) that couple different oscillation modes together (more details are discussed in \citealt{dewberry2023b}). The discrepancy in the dissipation away from resonances as noted by \cite{townsend2023} motivates us to take the approach described in Section \ref{sec:physics} of inverting Equation \ref{eq:Dklm} to find the Love number instead of using Equation \ref{eq:klmsum}.

In agreement with the inconsistency put forth by \cite{townsend2023}, we demonstrate in Figure \ref{fig:comparemethods} that the spectrum of $\mathrm{Im}\left[k_{2,2}^2\right]$ from Equation \ref{eq:klmsum} diverges from the results of Equations \ref{eq:D}--\ref{eq:Dklm} significantly away from resonance. In both calculations, we sum over the subset of f-modes and i-modes utilized in this work. However, Equation \ref{eq:klmsum} (blue dotted line) can yield unphysical results. For instance, it does not cross zero at $\omega_m=0$ with this set of modes, and the magnitude of the off-resonance dissipation is lower than what we find from our approach using the same viscosity. Figure \ref{fig:comparemethods} also verifies that our approach (red line) agrees very well with the results of a direct calculation (brown dots). Thus, the mode decomposition method can yield accurate results, but only if mode interaction cross terms due to dissipation (equation \ref{eq:D} in this work) are included when summing over modes.

\section{Conclusion}
We have modeled the frequency-dependent tidal response throughout the pre-main sequence and main sequence evolution of fully convective stars using realistic stellar models. Using a non-perturbative spectral method that fully resolves the effects of the Coriolis force to calculate fundamental and inertial mode oscillations of the star, we then compute the dissipation in terms of contributions from these normal modes. Our formalism accounts for the importance of viscous coupling between different modes in producing a reliable dissipation spectrum across tidal forcing frequency.

We performed our analysis for a fully-convective $0.2\, M_{\odot}$ M-dwarf at three different initial rotation periods, one of which ($P_{\rm rot,0}=1$ day) rotates very quickly relative to observed cluster stars, and two of which ($P_{\rm rot,0}=5$ and $10$ day) span a more common range of stellar rotation periods. As the star traverses the pre-main sequence and main sequence, we prescribe its spin evolution using the saturated magnetic braking model of \cite{Matt_2015}. The tidal response within this fully convective star is expected to be damped by a turbulent convective viscosity, which can be estimated from mixing-length theory as independent of Rossby number $\mathrm{Ro}$ (Equation \ref{eq:NRvisc}) or modified to account for the reduced efficiency of turbulent convective viscosity at high rotation rates (Equation \ref{eq:Rvisc}). Including the effects of rotation on viscosity decreases the magnitude of the viscosity in the stellar interior, leading to smaller equilibrium tidal dissipation but larger dissipation at resonances with stellar oscillation modes.
%Though we take the latter assumption for our fiducial models, we consider the effects of both approaches.

Using the formalism described in Section \ref{sec:physics}, we compute the spectrum of dynamical tidal dissipation across tidal forcing frequency $\omega_m$ throughout the stellar evolution. Peaks in the dissipation spectrum form at frequencies which are resonant with the star's inertial modes (i-modes) and fundamental modes (f-modes). For rapidly-rotating models $\Omega_s/\Omega_{\rm dyn} \gtrsim 0.6$, avoided crossings between f-modes and i-modes greatly enhance the dissipation at tidal forcing frequencies $|\omega_m| \leq 2\Omega_s$. As the star evolves, the mode frequencies change, allowing them to sweep through resonances with orbiting planets.

We integrate the semimajor axis evolution of planets causing tidal dissipation in the host star. We find that both inward and outward orbital migration can occur for Earth-mass planets at $P_{\rm orb,0} \lesssim 1.5$ day and Jupiter-mass planets at $P_{\rm orb,0}\lesssim 2.5$ day. In general, significant migration is induced by resonance locking, whereby the planetary orbit's tidal forcing frequency resonantly excites a stellar oscillation mode and continues to closely follow this mode frequency over time. This leads to enhanced dissipation during the resonance lock by several orders of magnitude. Due to orbital migration during the first few Gyr of the star's lifetime, we predict that the region of $P_{\rm orb}\lesssim 1$ day is likely to be cleared of small planets, and giant planets are made scarce out to $P_{\rm orb}\lesssim 3$ day around fully-convective M-dwarfs.

In the future, we hope to incorporate the feedback of the tidal torque on the stellar spin in our calculations, which we find to be significant for Jupiter-mass planets. This will contextualize our expectations for resonance locking of Jupiter-mass planets, as the spin-up or spin-down of the host star will affect the mode frequencies as a function of time. Another promising next step will be to apply the formalism established in this work to higher-mass, partially-convective stars. These stars will host gravito-inertial modes, which we anticipate will lead to a denser spectrum of dynamical tidal dissipation that can also drive resonance locking in planetary systems. These solar-type stars are representative of the majority of known exoplanet hosts and in particular the host star in the only planetary system whose orbit is known to decay, WASP-12b \citep{maciejewski2013, maciejewski2018, patra2017, yee2020}.

In addition, observations of exoplanet hosts in this mass range indicate that cooler stars with convective envelopes, which are often observed with spin that is aligned with their planetary orbits, may experience greater tidal dissipation than hotter stars with radiative envelopes, whose spin-orbit alignments trend larger \citep{albrecht2012,albrecht2021,winn2010,winn2015,spalding2022}. While we only considered aligned spin and orbits, our methods can easily be extended to planets with misaligned orbits to compute the tidal damping of obliquities. Studying the dynamical tide in low and high-mass stars will elucidate the puzzle of obliquity damping in these systems.

\section*{Acknowledgments}

This material is based upon work supported by the National Science Foundation Graduate Research Fellowship under Grant No. DGE‐1745301, and by NASA through grant 20-XRP20 2-0147.

\bibliography{bib}

\begin{thebibliography}{}
\expandafter\ifx\csname natexlab\endcsname\relax\def\natexlab#1{#1}\fi
\providecommand{\url}[1]{\href{#1}{#1}}

\bibitem[{{Albrecht} {et~al.}(2012){Albrecht}, {Winn}, {Johnson}, {Howard},
  {Marcy}, {Butler}, {Arriagada}, {Crane}, {Shectman}, {Thompson}, {Hirano},
  {Bakos}, \& {Hartman}}]{albrecht2012}
{Albrecht}, S., {Winn}, J.~N., {Johnson}, J.~A., {et~al.} 2012, \apj, 757, 18

\bibitem[{{Albrecht} {et~al.}(2021){Albrecht}, {Marcussen}, {Winn}, {Dawson},
  \& {Knudstrup}}]{albrecht2021}
{Albrecht}, S.~H., {Marcussen}, M.~L., {Winn}, J.~N., {Dawson}, R.~I., \&
  {Knudstrup}, E. 2021, \apjl, 916, L1

\bibitem[{{Bailey} \& {Goodman}(2019)}]{bailey2019}
{Bailey}, A., \& {Goodman}, J. 2019, \mnras, 482, 1872

\bibitem[{Barker(2020)}]{barker2020}
Barker, A.~J. 2020, Monthly Notices of the Royal Astronomical Society, 498,
  2270.
\newblock \url{https://doi.org/10.1093/mnras/staa2405}

\bibitem[{{Barker} {et~al.}(2014){Barker}, {Dempsey}, \&
  {Lithwick}}]{barker2014}
{Barker}, A.~J., {Dempsey}, A.~M., \& {Lithwick}, Y. 2014, \apj, 791, 13

\bibitem[{{Bolmont} {et~al.}(2017){Bolmont}, {Gallet}, {Mathis}, {Charbonnel},
  {Amard}, \& {Alibert}}]{bolmont2017}
{Bolmont}, E., {Gallet}, F., {Mathis}, S., {et~al.} 2017, \aap, 604, A113

\bibitem[{{Bolmont} \& {Mathis}(2016)}]{bolmont2016}
{Bolmont}, E., \& {Mathis}, S. 2016, Celestial Mechanics and Dynamical
  Astronomy, 126, 275

\bibitem[{{Bonomo} {et~al.}(2017){Bonomo}, {Desidera}, {Benatti}, {Borsa},
  {Crespi}, {Damasso}, {Lanza}, {Sozzetti}, {Lodato}, {Marzari}, {Boccato},
  {Claudi}, {Cosentino}, {Covino}, {Gratton}, {Maggio}, {Micela}, {Molinari},
  {Pagano}, {Piotto}, {Poretti}, {Smareglia}, {Affer}, {Biazzo}, {Bignamini},
  {Esposito}, {Giacobbe}, {H{\'e}brard}, {Malavolta}, {Maldonado}, {Mancini},
  {Martinez Fiorenzano}, {Masiero}, {Nascimbeni}, {Pedani}, {Rainer}, \&
  {Scandariato}}]{bonomo2017}
{Bonomo}, A.~S., {Desidera}, S., {Benatti}, S., {et~al.} 2017, \aap, 602, A107

\bibitem[{{Braviner} \& {Ogilvie}(2015)}]{braviner2015}
{Braviner}, H.~J., \& {Ogilvie}, G.~I. 2015, \mnras, 447, 1141

\bibitem[{{Burkart} {et~al.}(2012){Burkart}, {Quataert}, {Arras}, \&
  {Weinberg}}]{Burkart2012}
{Burkart}, J., {Quataert}, E., {Arras}, P., \& {Weinberg}, N.~N. 2012, \mnras,
  421, 983

\bibitem[{{Dawson} \& {Johnson}(2018)}]{dawson2018}
{Dawson}, R.~I., \& {Johnson}, J.~A. 2018, \araa, 56, 175

\bibitem[{{Dewberry}(2023)}]{Dewberry2023}
{Dewberry}, J.~W. 2023, \mnras, 521, 5991

\bibitem[{{Dewberry} \& {Lai}(2022)}]{Dewberry2022}
{Dewberry}, J.~W., \& {Lai}, D. 2022, \apj, 925, 124

\bibitem[{Dewberry {et~al.}(2021)Dewberry, Mankovich, Fuller, Lai, \&
  Xu}]{Dewberry2021}
Dewberry, J.~W., Mankovich, C.~R., Fuller, J., Lai, D., \& Xu, W. 2021, The
  Planetary Science Journal, 2, 198.
\newblock \url{https://dx.doi.org/10.3847/PSJ/ac0e2a}

\bibitem[{{Dewberry} \& {Wu}(2023)}]{dewberry2023b}
{Dewberry}, J.~W., \& {Wu}, S.~C. 2023, arXiv e-prints, arXiv:2309.11502

\bibitem[{{Dressing} \& {Charbonneau}(2013)}]{dressing2013}
{Dressing}, C.~D., \& {Charbonneau}, D. 2013, \apj, 767, 95

\bibitem[{{Duguid} {et~al.}(2020{\natexlab{a}}){Duguid}, {Barker}, \&
  {Jones}}]{duguid2020a}
{Duguid}, C.~D., {Barker}, A.~J., \& {Jones}, C.~A. 2020{\natexlab{a}}, \mnras,
  497, 3400

\bibitem[{{Duguid} {et~al.}(2020{\natexlab{b}}){Duguid}, {Barker}, \&
  {Jones}}]{duguid2020b}
---. 2020{\natexlab{b}}, \mnras, 491, 923

\bibitem[{{Fuller}(2017)}]{fuller2017}
{Fuller}, J. 2017, \mnras, 472, 1538

\bibitem[{{Fuller} \& {Lai}(2012)}]{fuller2012}
{Fuller}, J., \& {Lai}, D. 2012, \mnras, 420, 3126

\bibitem[{{Fuller} {et~al.}(2016){Fuller}, {Luan}, \& {Quataert}}]{fuller2016}
{Fuller}, J., {Luan}, J., \& {Quataert}, E. 2016, \mnras, 458, 3867

\bibitem[{{Gallet} {et~al.}(2017){Gallet}, {Bolmont}, {Mathis}, {Charbonnel},
  \& {Amard}}]{gallet2017}
{Gallet}, F., {Bolmont}, E., {Mathis}, S., {Charbonnel}, C., \& {Amard}, L.
  2017, \aap, 604, A112

\bibitem[{{Gillon} {et~al.}(2017){Gillon}, {Triaud}, {Demory}, {Jehin}, {Agol},
  {Deck}, {Lederer}, {de Wit}, {Burdanov}, {Ingalls}, {Bolmont}, {Leconte},
  {Raymond}, {Selsis}, {Turbet}, {Barkaoui}, {Burgasser}, {Burleigh}, {Carey},
  {Chaushev}, {Copperwheat}, {Delrez}, {Fernandes}, {Holdsworth}, {Kotze}, {Van
  Grootel}, {Almleaky}, {Benkhaldoun}, {Magain}, \&
  {Queloz}}]{2017Natur.542..456G}
{Gillon}, M., {Triaud}, A. H.~M.~J., {Demory}, B.-O., {et~al.} 2017, \nat, 542,
  456

\bibitem[{{Goldreich} \& {Keeley}(1977)}]{goldreich1977}
{Goldreich}, P., \& {Keeley}, D.~A. 1977, \apj, 211, 934

\bibitem[{{Goodman} \& {Lackner}(2009)}]{goodman2009}
{Goodman}, J., \& {Lackner}, C. 2009, \apj, 696, 2054

\bibitem[{{Goodman} \& {Oh}(1997)}]{goodman1997}
{Goodman}, J., \& {Oh}, S.~P. 1997, \apj, 486, 403

\bibitem[{{Hsu} {et~al.}(2020){Hsu}, {Ford}, \& {Terrien}}]{hsu2020}
{Hsu}, D.~C., {Ford}, E.~B., \& {Terrien}, R. 2020, \mnras, 498, 2249

\bibitem[{{Ipser} \& {Lindblom}(1991)}]{ipser1991}
{Ipser}, J.~R., \& {Lindblom}, L. 1991, \apj, 373, 213

\bibitem[{{Kumar} {et~al.}(1995){Kumar}, {Ao}, \& {Quataert}}]{Kumar1995}
{Kumar}, P., {Ao}, C.~O., \& {Quataert}, E.~J. 1995, \apj, 449, 294

\bibitem[{{Lai}(1997)}]{lai1997}
{Lai}, D. 1997, \apj, 490, 847

\bibitem[{{Lai} \& {Wu}(2006)}]{Lai2006}
{Lai}, D., \& {Wu}, Y. 2006, \prd, 74, 024007

\bibitem[{{Lee} \& {Chiang}(2017)}]{lee2017}
{Lee}, E.~J., \& {Chiang}, E. 2017, \apj, 842, 40

\bibitem[{{Maciejewski} {et~al.}(2013){Maciejewski}, {Dimitrov}, {Seeliger},
  {Raetz}, {Bukowiecki}, {Kitze}, {Errmann}, {Nowak}, {Niedzielski}, {Popov},
  {Marka}, {Go{\'z}dziewski}, {Neuh{\"a}user}, {Ohlert}, {Hinse}, {Lee}, {Lee},
  {Yoon}, {Berndt}, {Gilbert}, {Ginski}, {Hohle}, {Mugrauer}, {R{\"o}ll},
  {Schmidt}, {Tetzlaff}, {Mancini}, {Southworth}, {Dall'Ora}, {Ciceri},
  {Zambelli}, {Corfini}, {Takahashi}, {Tachihara}, {Benk{\H{o}}},
  {S{\'a}rneczky}, {Szabo}, {Varga}, {Va{\v{n}}ko}, {Joshi}, \&
  {Chen}}]{maciejewski2013}
{Maciejewski}, G., {Dimitrov}, D., {Seeliger}, M., {et~al.} 2013, \aap, 551,
  A108

\bibitem[{{Maciejewski} {et~al.}(2018){Maciejewski}, {Fern{\'a}ndez},
  {Aceituno}, {Mart{\'\i}n-Ruiz}, {Ohlert}, {Dimitrov}, {Szyszka}, {von Essen},
  {Mugrauer}, {Bischoff}, {Michel}, {Mallonn}, {Stangret}, \&
  {Mo{\'z}dzierski}}]{maciejewski2018}
{Maciejewski}, G., {Fern{\'a}ndez}, M., {Aceituno}, F., {et~al.} 2018, \actaa,
  68, 371

\bibitem[{{Mathis} {et~al.}(2016){Mathis}, {Auclair-Desrotour}, {Guenel},
  {Gallet}, \& {Le Poncin-Lafitte}}]{mathis2016}
{Mathis}, S., {Auclair-Desrotour}, P., {Guenel}, M., {Gallet}, F., \& {Le
  Poncin-Lafitte}, C. 2016, \aap, 592, A33

\bibitem[{Matt {et~al.}(2015)Matt, Brun, Baraffe, Bouvier, \&
  Chabrier}]{Matt_2015}
Matt, S.~P., Brun, A.~S., Baraffe, I., Bouvier, J., \& Chabrier, G. 2015, The
  Astrophysical Journal Letters, 799, L23.
\newblock \url{https://dx.doi.org/10.1088/2041-8205/799/2/L23}

\bibitem[{{Ment} \& {Charbonneau}(2023)}]{ment2023}
{Ment}, K., \& {Charbonneau}, D. 2023, \aj, 165, 265

\bibitem[{{Ment} {et~al.}(2021){Ment}, {Irwin}, {Charbonneau}, {Winters},
  {Medina}, {Cloutier}, {D{\'\i}az}, {Jenkins}, {Ziegler}, {Law}, {Mann},
  {Ricker}, {Vanderspek}, {Latham}, {Seager}, {Winn}, {Jenkins}, {Goeke},
  {Levine}, {Rojas-Ayala}, {Rowden}, {Ting}, \& {Twicken}}]{ment2021}
{Ment}, K., {Irwin}, J., {Charbonneau}, D., {et~al.} 2021, \aj, 161, 23

\bibitem[{Mulders {et~al.}(2015)Mulders, Pascucci, \& Apai}]{Mulders_2015}
Mulders, G.~D., Pascucci, I., \& Apai, D. 2015, The Astrophysical Journal, 798,
  112.
\newblock \url{https://dx.doi.org/10.1088/0004-637X/798/2/112}

\bibitem[{{Ogilvie}(2013)}]{ogilvie2013}
{Ogilvie}, G.~I. 2013, \mnras, 429, 613

\bibitem[{{Ogilvie}(2014)}]{ogilvie2014}
---. 2014, \araa, 52, 171

\bibitem[{{Ogilvie} \& {Lesur}(2012)}]{ogilvie2012}
{Ogilvie}, G.~I., \& {Lesur}, G. 2012, \mnras, 422, 1975

\bibitem[{{Ogilvie} \& {Lin}(2007)}]{ogilvie2007}
{Ogilvie}, G.~I., \& {Lin}, D.~N.~C. 2007, \apj, 661, 1180

\bibitem[{{Patra} {et~al.}(2017){Patra}, {Winn}, {Holman}, {Yu}, {Deming}, \&
  {Dai}}]{patra2017}
{Patra}, K.~C., {Winn}, J.~N., {Holman}, M.~J., {et~al.} 2017, \aj, 154, 4

\bibitem[{{Paxton} {et~al.}(2011){Paxton}, {Bildsten}, {Dotter}, {Herwig},
  {Lesaffre}, \& {Timmes}}]{mesa2011}
{Paxton}, B., {Bildsten}, L., {Dotter}, A., {et~al.} 2011, \apjs, 192, 3

\bibitem[{{Paxton} {et~al.}(2013){Paxton}, {Cantiello}, {Arras}, {Bildsten},
  {Brown}, {Dotter}, {Mankovich}, {Montgomery}, {Stello}, {Timmes}, \&
  {Townsend}}]{mesa2013}
{Paxton}, B., {Cantiello}, M., {Arras}, P., {et~al.} 2013, \apjs, 208, 4

\bibitem[{{Paxton} {et~al.}(2015){Paxton}, {Marchant}, {Schwab}, {Bauer},
  {Bildsten}, {Cantiello}, {Dessart}, {Farmer}, {Hu}, {Langer}, {Townsend},
  {Townsley}, \& {Timmes}}]{mesa2015}
{Paxton}, B., {Marchant}, P., {Schwab}, J., {et~al.} 2015, \apjs, 220, 15

\bibitem[{{Paxton} {et~al.}(2018){Paxton}, {Schwab}, {Bauer}, {Bildsten},
  {Blinnikov}, {Duffell}, {Farmer}, {Goldberg}, {Marchant}, {Sorokina},
  {Thoul}, {Townsend}, \& {Timmes}}]{mesa2018}
{Paxton}, B., {Schwab}, J., {Bauer}, E.~B., {et~al.} 2018, \apjs, 234, 34

\bibitem[{{Paxton} {et~al.}(2019){Paxton}, {Smolec}, {Schwab}, {Gautschy},
  {Bildsten}, {Cantiello}, {Dotter}, {Farmer}, {Goldberg}, {Jermyn}, {Kanbur},
  {Marchant}, {Thoul}, {Townsend}, {Wolf}, {Zhang}, \& {Timmes}}]{mesa2019}
{Paxton}, B., {Smolec}, R., {Schwab}, J., {et~al.} 2019, \apjs, 243, 10

\bibitem[{{Penev} {et~al.}(2009){Penev}, {Barranco}, \& {Sasselov}}]{penev2009}
{Penev}, K., {Barranco}, J., \& {Sasselov}, D. 2009, \apj, 705, 285

\bibitem[{{Penev} {et~al.}(2007){Penev}, {Sasselov}, {Robinson}, \&
  {Demarque}}]{penev2007}
{Penev}, K., {Sasselov}, D., {Robinson}, F., \& {Demarque}, P. 2007, \apj, 655,
  1166

\bibitem[{{Press} \& {Teukolsky}(1977)}]{press1977}
{Press}, W.~H., \& {Teukolsky}, S.~A. 1977, \apj, 213, 183

\bibitem[{{Rodr{\'\i}guez Mart{\'\i}nez} {et~al.}(2023){Rodr{\'\i}guez
  Mart{\'\i}nez}, {Martin}, {Gaudi}, {Schulze}, {Asnodkar}, {Boley}, \&
  {Ballard}}]{rodriguez2023}
{Rodr{\'\i}guez Mart{\'\i}nez}, R., {Martin}, D.~V., {Gaudi}, B.~S., {et~al.}
  2023, \aj, 166, 137

\bibitem[{{Schenk} {et~al.}(2001){Schenk}, {Arras}, {Flanagan}, {Teukolsky}, \&
  {Wasserman}}]{Schenk2001}
{Schenk}, A.~K., {Arras}, P., {Flanagan}, {\'E}.~{\'E}., {Teukolsky}, S.~A., \&
  {Wasserman}, I. 2001, \prd, 65, 024001

\bibitem[{{Spalding} \& {Winn}(2022)}]{spalding2022}
{Spalding}, C., \& {Winn}, J.~N. 2022, \apj, 927, 22

\bibitem[{{Stefansson} {et~al.}(2020){Stefansson}, {Mahadevan}, {Maney},
  {Ninan}, {Robertson}, {Rajagopal}, {Haase}, {Allen}, {Ford}, {Winn},
  {Wolfgang}, {Dawson}, {Wisniewski}, {Bender}, {Ca{\~n}as}, {Cochran},
  {Diddams}, {Fredrick}, {Halverson}, {Hearty}, {Hebb}, {Kanodia}, {Levi},
  {Metcalf}, {Monson}, {Ramsey}, {Roy}, {Schwab}, {Terrien}, \&
  {Wright}}]{stefansson2020}
{Stefansson}, G., {Mahadevan}, S., {Maney}, M., {et~al.} 2020, \aj, 160, 192

\bibitem[{{Stevenson}(1979)}]{stevenson1979}
{Stevenson}, D.~J. 1979, Geophysical and Astrophysical Fluid Dynamics, 12, 139

\bibitem[{{Tamburo} {et~al.}(2023){Tamburo}, {Muirhead}, \&
  {Dressing}}]{tamburo2023}
{Tamburo}, P., {Muirhead}, P.~S., \& {Dressing}, C.~D. 2023, \aj, 165, 251

\bibitem[{{Terquem}(2021)}]{terquem2021}
{Terquem}, C. 2021, \mnras, 503, 5789

\bibitem[{{Terquem}(2023)}]{terquem2023}
---. 2023, \mnras, 525, 508

\bibitem[{{Thao} {et~al.}(2020){Thao}, {Mann}, {Johnson}, {Newton}, {Guo},
  {Kain}, {Rizzuto}, {Charbonneau}, {Dalba}, {Gaidos}, {Irwin}, \&
  {Kraus}}]{thao2020}
{Thao}, P.~C., {Mann}, A.~W., {Johnson}, M.~C., {et~al.} 2020, \aj, 159, 32

\bibitem[{{Townsend} \& {Sun}(2023)}]{townsend2023}
{Townsend}, R.~H.~D., \& {Sun}, M. 2023, \apj, 953, 48

\bibitem[{{Vick} \& {Lai}(2020)}]{vick2020}
{Vick}, M., \& {Lai}, D. 2020, \mnras, 496, 3767

\bibitem[{{Weinberg} {et~al.}(2012){Weinberg}, {Arras}, {Quataert}, \&
  {Burkart}}]{weinberg2012}
{Weinberg}, N.~N., {Arras}, P., {Quataert}, E., \& {Burkart}, J. 2012, \apj,
  751, 136

\bibitem[{{Weinberg} {et~al.}(2017){Weinberg}, {Sun}, {Arras}, \&
  {Essick}}]{weinberg2017}
{Weinberg}, N.~N., {Sun}, M., {Arras}, P., \& {Essick}, R. 2017, \apjl, 849,
  L11

\bibitem[{{Winn} {et~al.}(2010){Winn}, {Fabrycky}, {Albrecht}, \&
  {Johnson}}]{winn2010}
{Winn}, J.~N., {Fabrycky}, D., {Albrecht}, S., \& {Johnson}, J.~A. 2010, \apjl,
  718, L145

\bibitem[{{Winn} \& {Fabrycky}(2015)}]{winn2015}
{Winn}, J.~N., \& {Fabrycky}, D.~C. 2015, \araa, 53, 409

\bibitem[{{Wirth} {et~al.}(2021){Wirth}, {Zhou}, {Quinn}, {Mann}, {Bouma},
  {Latham}, {Teske}, {Wang}, {Shectman}, {Butler}, \& {Crane}}]{wirth2021}
{Wirth}, C.~P., {Zhou}, G., {Quinn}, S.~N., {et~al.} 2021, \apjl, 917, L34

\bibitem[{{Witte} \& {Savonije}(1999)}]{witte1999}
{Witte}, M.~G., \& {Savonije}, G.~J. 1999, \aap, 350, 129

\bibitem[{{Wu}(2005{\natexlab{a}})}]{wu2005a}
{Wu}, Y. 2005{\natexlab{a}}, \apj, 635, 674

\bibitem[{{Wu}(2005{\natexlab{b}})}]{wu2005b}
---. 2005{\natexlab{b}}, \apj, 635, 688

\bibitem[{{Yee} {et~al.}(2020){Yee}, {Winn}, {Knutson}, {Patra},
  {Vissapragada}, {Zhang}, {Holman}, {Shporer}, \& {Wright}}]{yee2020}
{Yee}, S.~W., {Winn}, J.~N., {Knutson}, H.~A., {et~al.} 2020, \apjl, 888, L5

\bibitem[{{Zahn}(1966{\natexlab{a}})}]{zahn1966a}
{Zahn}, J.~P. 1966{\natexlab{a}}, Annales d'Astrophysique, 29, 313

\bibitem[{{Zahn}(1966{\natexlab{b}})}]{zahn1966b}
---. 1966{\natexlab{b}}, Annales d'Astrophysique, 29, 489

\end{thebibliography}
\end{document}